\documentclass[twocolumn]{aastex63}

\usepackage{verbatim}
\usepackage{graphicx,url}
\usepackage{amsmath,amssymb}
\usepackage{hyperref}
\hypersetup{colorlinks,
linkcolor=blue, 
urlcolor=magenta, 
anchorcolor=blue,
citecolor=blue
}
\usepackage{subfigure}

\begin{document}

\title{Delensing of Cosmic Microwave Background Polarization with machine learning}


\author{Ye-Peng Yan}
\affil{Department of Astronomy, Beijing Normal University, Beijing 100875, China; xiajq@bnu.edu.cn}

\author{Guo-Jian Wang}
\affil{School of Chemistry and Physics, University of KwaZulu-Natal, Westville Campus, Private Bag X54001 Durban, 4000, South Africa}
\affil{NAOC-UKZN Computational Astrophysics Centre (NUCAC), University of KwaZulu-Natal, Durban, 4000, South Africa}

\author{Si-Yu Li}
\affil{Key Laboratory of Particle Astrophysics, Institute of High Energy Physics, Chinese Academy of Science, P. O. Box 918-3 Beijing 100049, People’s Republic of China}

\author{Jun-Qing Xia}
\affil{Department of Astronomy, Beijing Normal University, Beijing 100875, China; xiajq@bnu.edu.cn}

\begin{abstract}
Primordial B-mode detection is one of the main goals of next-generation cosmic microwave background (CMB) experiments. Primordial B-modes are a unique signature of primordial gravitational waves (PGWs). However, the gravitational interaction of CMB photons with large-scale structures will distort the primordial E modes, adding a lensing B-mode component to the primordial B-mode signal. Removing the lensing effect (`delensing') from observed CMB polarization maps will be necessary to improve the constraint of PGWs and obtain a primordial E-mode signal. Here, we introduce a deep convolutional neural network model named multi-input multi-output U-net (MIMO-UNet) to perform CMB delensing. The networks are trained on simulated CMB maps with size $20^{\circ} \times 20^{\circ}$. We first use MIMO-UNet to reconstruct the unlensing CMB polarization ($Q$ and $U$) maps from observed CMB maps. The recovered E-mode power spectrum exhibits excellent agreement with the primordial EE power spectrum. The recovery of the primordial B-mode power spectrum for noise levels of 0, 1, and 2 $\mu$K-arcmin is greater than 98\%  at the angular scale of $\ell<150$. We additionally reconstruct the lensing  B map from observed CMB maps. The recovery of the lensing  B-mode power spectrum is greater than roughly 99\% at the scales of $\ell>200$. We delens observed B-mode power spectrum by subtracting reconstructed lensing B-mode spectrum. The recovery of tensor B-mode power spectrum for noise levels of 0, 1, 2 $\mu$K-arcmin is greater than 98 \% at the angular scales of $\ell<120$. Even at  $\ell=160$, the recovery of tensor B-mode power spectrum is still around  71 \%.
\end{abstract}


\section{INTRODUCTION} 
Inflationary models generically predict the existence of a stochastic background of gravitational waves (i.e., primordial gravitational waves (PGWs)) \citep{Kamionkowski2016}. The tensor-to-scalar ratio $r$, which parameterize the PGW's amplitude, is related to the energy scale at which inflation occurred. Because of this, it is anticipated that the detection of PGWs will reveal the physics of the universe's very early stages. Fortunately, these PGWs imprint a unique signature on the polarized anisotropies of the cosmic microwave background (CMB) \citep{Kamionkowski1997,Zaldarriaga1997}  in the form of curl-like patterns B-modes, which makes it possible to extract a signal of the PGW from B-modes of the CMB polarization. To constrain the tensor-to-scalar ratio with high precision \citep{Abazajian2022}, several next-generation CMB experiments with  multi-frequency coverage and very high  sensitivity, such as the CMB-S4 project \citep{Abazajian2019},  LiteBIRD satellite \citep{Hazumi2019}, and AliCPT \citep{Li2017}, have been proposed or are under construction. However, PGWs are not the only source of B-modes. Gravitational lensing of the CMB is a serious obstacle to the search for PGWs.  Gravitational lensing of the CMB arises from the deflection of CMB photons as they pass through the matter distribution between the surface of the last scattering and us, which results in a subtle remapping of the temperature and polarization anisotropies \citep{Lewis2006}. Gravitational lensing can convert a small part of E-modes polarization into B-modes polarization, which acts as a source of confusion in searches for PGWs. Furthermore, lensing also smooths the acoustic peaks of the CMB power spectra and produces non-stationary statistics of CMB fluctuations \citep{Zaldarriaga1998,Lewis2001,Hotinli2022}.  In order to improve constraints on the tensor-to-scalar ratio, we must remove the effects of lensing (delensing) from observed CMB.  Additionally, delensing of observed CMB maps can sharpen acoustic peaks of T and E modes, and tighten cosmological parameter constraints \citep{Green2017}.

CMB delensing methods have been investigated  for many years \citep{Knox2002,Kesden2002,Seljak2004,Smith2012,Larsen2016,Carron2017,Millea2019,Millea2020,SushovanChandra2021}.  Internal delensing methods require the subtraction of a template of the lensing B-mode constructed from observed E-modes and a tracer of the mass distribution that lensed the CMB.  External delensing uses external mass tracers which correlate
with the CMB lensing signal, such as the cosmic infrared background (CIB) \citep{Simard2015,Sherwin2015,Yu2017}, radio and optical galaxies \citep{Namikawa2016,Manzotti2018}, and intensity maps of high-redshift line
emissions\citep{Sigurdson2005,Karkare2019}.  Some of these delensing methods have already been demonstrated on data \citep{Carron2017,Manzotti2017,PlanckCollaboration2020,Adachi2020,Han2021,BICEP/KeckCollaboration2021}, and have been used to constraint tensor-to-scalar ratio. The current best constraints on the tensor-to-scalar ratio are $r<0.06$ at $95\%$ confidence obtained from the BICEP2/Keck Array \citep{BICEP2Collaboration2018} or $r<0.036$ at $95\%$ \citep{Ade2021} based on the BICEP2, Keck Array, and BICEP3 CMB polarization experiments.

\begin{figure*}
\begin{center}
	\includegraphics[width=1\hsize]{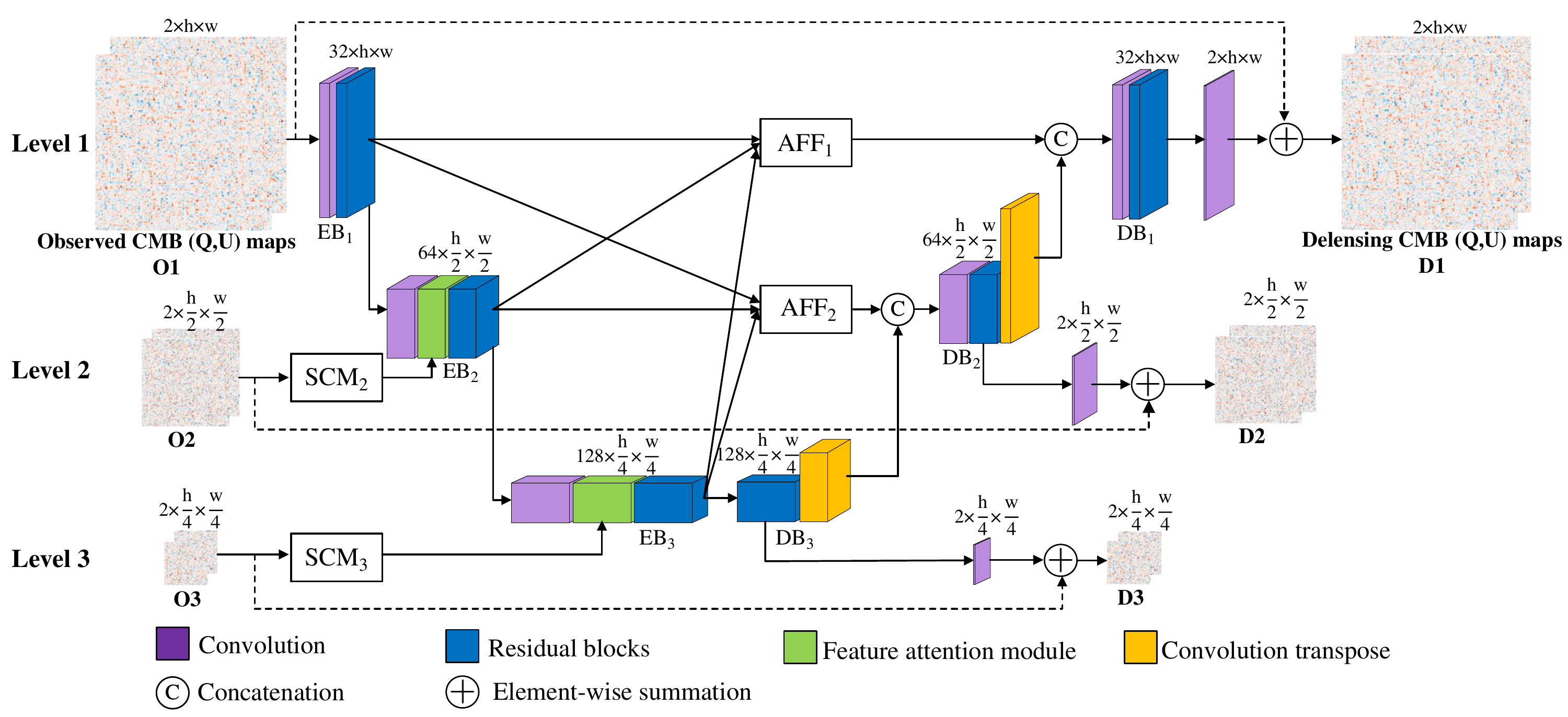}
\end{center}
\vspace{-0.4cm}
\caption{The architecture of the MIMO-UNet. The input images are observed CMB (Q and U) maps, which is denoted by $O_1$. The $O_2$ and $O_3$ are down-sample of $O1$ using the interpolation method. The sizes (${\rm (c\times h \times w)=( channel \times height \times width})$) of each module's feature images are shown in figure. The input images ($O_1, O_2, O_3$) pass through the convolution layers of related levels, and finally the three output images ($D_1$, $D_2$, $D_3$) are obtained.}
\label{figure_net}
\end{figure*}

\begin{figure}
\begin{center}
	\includegraphics[width=1\hsize]{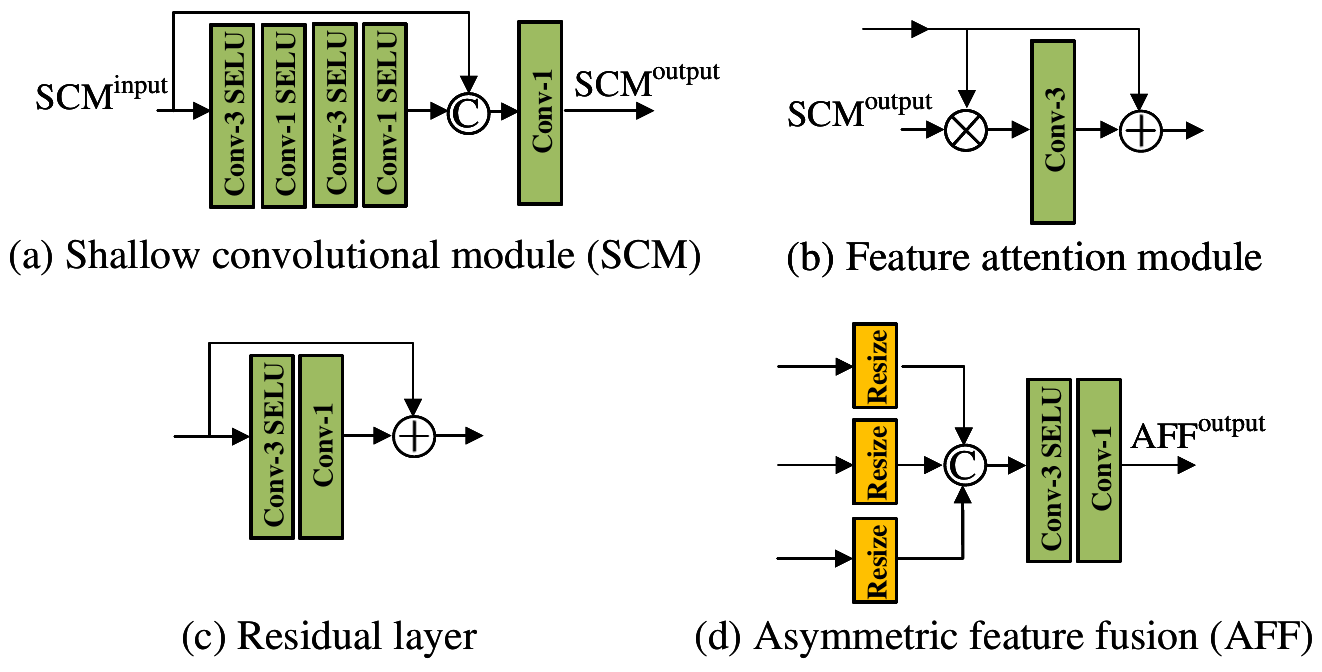}
\end{center}
\vspace{-0.4cm}
\caption{The structures of sub-modules of the MIMO-UNet. Conv-3 and Conv-1 denote convolutional layers with $3\times3$ and $1\times1$ filters, respectively. SELU and $\bigotimes$ represent  Scaled Exponential Linear Unit (SELU) function and multiplication, respectively. }
\label{figure_net_comp}
\end{figure}
With the remarkable progress of computer science in recent years, machine learning (ML) has demonstrated exceptional capabilities in the image processing field, and it has been progressively used to CMB data processing. For example, the methods of ML have been applied successfully to the CMB component separation \citep{Petroff2020,Wang2022,Casas2022,Yan2023}. In addition, Refs. \citep{Caldeira2019, Guzman2021, Li2022, Heinrich2022,Guzman2022}   also employ the ML approach for the reconstruction of CMB lensing and cosmic polarization rotation maps. \cite{Caldeira2019} also uses ML method to derive delensing E map for observed CMB Q/U maps. However, the delensing EE power spectrum obviously deviates from true EE at the angular scales of $\ell>1000$.  \cite{Guzman2021} and \cite{Heinrich2022} improve the performance in delensing E map, but the delensing EE power spectrum also deviates from true EE with $\ell$ increases at the angular scales of $\ell>1800$. As an improvement and development, in this work, we introduce a convolutional neural network (CNN) to perform CMB delensing.

This paper is organized as follows. In Section \ref{Methodology}, we describe our network model as well as the simulated data sets. In Section \ref{delensing_result}, we present the results of reconstructing the unlensing CMB ($Q$ and $U$) maps. In Section \ref{Reconstruct_lensing}, we present the results of reconstructing the lensing B map. In Section \ref{DISCUSSIONS} discussions are presented. Finally, we conclude in Section \ref{CONCLUSIONS}.

\section{Methodology}
\label{Methodology}
The deep learning architecture we use in this work, as well as the details of the data pipeline and network training, will all be covered in this section.

\subsection{Network architecture}

Machine learning has demonstrated exceptional capabilities in the field of image processing in recent years.  The convolutional neural network (CNN) is the most widely used type of feed-forward neural network in the tasks of image processing \citep{Mehta2019}. The CNN's core structure is the convolutional layer, which computes the convolution of the input image with a bank of filters (kernels) whose values are parameters to be learned. Each convolutional layer accepts the image from the former layer as an input, calculate the convolution of the input image, and finally gives output to the next layer after  applying a nonlinear activation function. The convolutional layer generally has three hyper-parameters: the number of output channels (the number of convolutional filters), stride, and amount of zero padding. The output size of a convolutional layer can be controlled by setting these three hyper-parameters. Here, the stride is defined as the distance in pixels between the centers of adjacent filters. 

In this work, we use the multi-input multi-output U-net (MIMO-UNet) \citep{Cho2021}, an excellent network architecture for image deblurring tasks, to perform the CMB delensing. The MIMO-UNet takes multi-scale input images to ease the difficulty of training. As shown in Figure \ref{figure_net}, MIMO-UNet model consists of three levels. To obtain the multi-scale images ($O_2$, $O_3$), we first perform down sampling for input image ($O_1$) using the interpolation method, then take three scale images ($O_1$, $O_2$, $O_3$) as an input of corresponding levels. 

In the first level, we use ${\rm EB_1}$ block, which consists of a  convolutional layer and a residual block, to extract the features from the input image $O_1$. This convolution layer is used to increase the number of channel, and the residual block is used to extract the features. The residual block consists of many residual layers as shown in panel (c) of Figure \ref{figure_net_comp}, and we set the number of residual layers to 16 for a residual block in this work. Then we use a shallow convolutional module (SCM) as shown in panel (a) to extract the features from downsampled images ($O_2$ and $O_3$). The SCM has two stacks of $3\times3$ and $1\times1$ convolutional layers. We concatenate the features from the last $1\times1$ layer with the input of SCM, and further refine the concatenated features using an additional $1\times1$ convolutional layer. The output of ${\rm EB_1}$ reduces the size by half through a convolutional layer and is fed to the feature attention module (FAM) as shown in panel (b) of Figure \ref{figure_net_comp}, and the FAM performs the element-wise multiplied of this input and output of ${\rm SCM_2}$, and then the multiplied features are passed through a $3\times 3$ convolutional layers. The output of FAM is fed to residual block to further refine features. Finally, the output of ${\rm EB_2}$ can be obtained. Similarly, the output of ${\rm EB_2}$ and output of ${\rm SCM_2}$ are fed to ${\rm EB_3}$. 

The outputs of ${\rm EB_2}$ and ${\rm EB_3}$ perform the up sample using the interpolation method, and are fed to ${\rm AFF_1}$ together with the output of ${\rm EB_1}$. The asymmetric feature fusion (AFF) module as shown in panel (d) of Figure \ref{figure_net_comp} allows information flow from different scales feature maps. Similarly, The outputs of ${\rm EB_2}$, ${\rm EB_3}$, and ${\rm EB_1}$ are also fed to ${\rm AFF_2}$. The output of the AFF is delivered to its corresponding DB. ${\rm DB_3}$ consists of a residual block and convolution transpose layer, which is used to double the sizes of the feature maps. The output of ${\rm DB_3}$ and ${\rm AFF_2}$ be concatenated and fed to ${\rm DB_2}$. Finally, the output of ${\rm DB_2}$ and ${\rm AFF_1}$ be concatenated and fed to ${\rm DB_1}$, and the delensing CMB Q and U maps can be obtained. In addition, the second-level and third-level DBs also output images before convolution transpose layer. The MIMO-UNet model is built in \texttt{PyTorch}\footnote{\url{https://pytorch.org/}} environment, which is an open-source optimized tensor library for deep learning.

The network model consists of a stack of non-linear parameters, and these parameters can be optimized by minimizing a loss function when the network is trained on a training set. We assume that the training set has $S$ pairs of samples $\{x_i, y_i\}_{i=1}^S$. Here, $x$ is the observed CMB Q and U maps, and $y$ represents the corresponding ground truth of CMB maps. Our loss function defines as 
\begin{align}
	\mathcal{L} = \sum^{3}_{k=1} (\mathcal{L}_{{\rm LAD},k} + \beta \mathcal{L}_{{\rm FFT},k}),
\end{align}
where $\mathcal{L}_{\rm LAD}$ is the least absolute deviation (LAD, also called L1 loss), $\mathcal{L}_{\rm FFT}$ is the Fourier space loss, $k$ is the number of levels of network, and $\beta$ is a coefficient representing the contribution of $\mathcal{L}_{\rm FFT}$ to the total loss, and we set $\beta=1$ throughout the paper. The L1 loss has form
\begin{align}
	\mathcal{L}_{\rm LAD}&=\frac{1}{N}\sum_{n=1}^{N}\left[ \frac{1}{WH}\sum_{w=1}^{W}\sum_{h=1}^{H}(|I^{n}_{w,h}-y^{n}_{w,h}|)\right],
\end{align} 
where $N$ is the batch size, $H$ ($W$) is the height (width) of the images in pixels, and $I = f(x)$ ($f(\cdot)$ is the network model) is the predicted image. The form of $\mathcal{L}_{\rm FFT}$ is defined as 
\begin{align}
	\mathcal{L}_{\rm FFT}&=\frac{1}{N}\sum_{n=1}^{N}\left[ \frac{1}{WH}\sum_{w=1}^{W}\sum_{h=1}^{H}(|A_{\rm F}(I^{n}_{w,h})-A_{\rm F}(y^{n}_{w,h})|)\right],
\end{align}
where $A_{\rm F}$ is the amplitude of FFT, which has the form of 
\begin{align}
	A_{\rm F}(I)& = \sqrt{Re[{\rm FFT}(I)]^2+Im[{\rm FFT}(I)]^2},
\end{align}
where $Re[\cdot]$ and $Im[\cdot]$ denote the real and imaginary part, respectively.
\begin{figure}
\begin{center}
	\includegraphics[width=1\hsize]{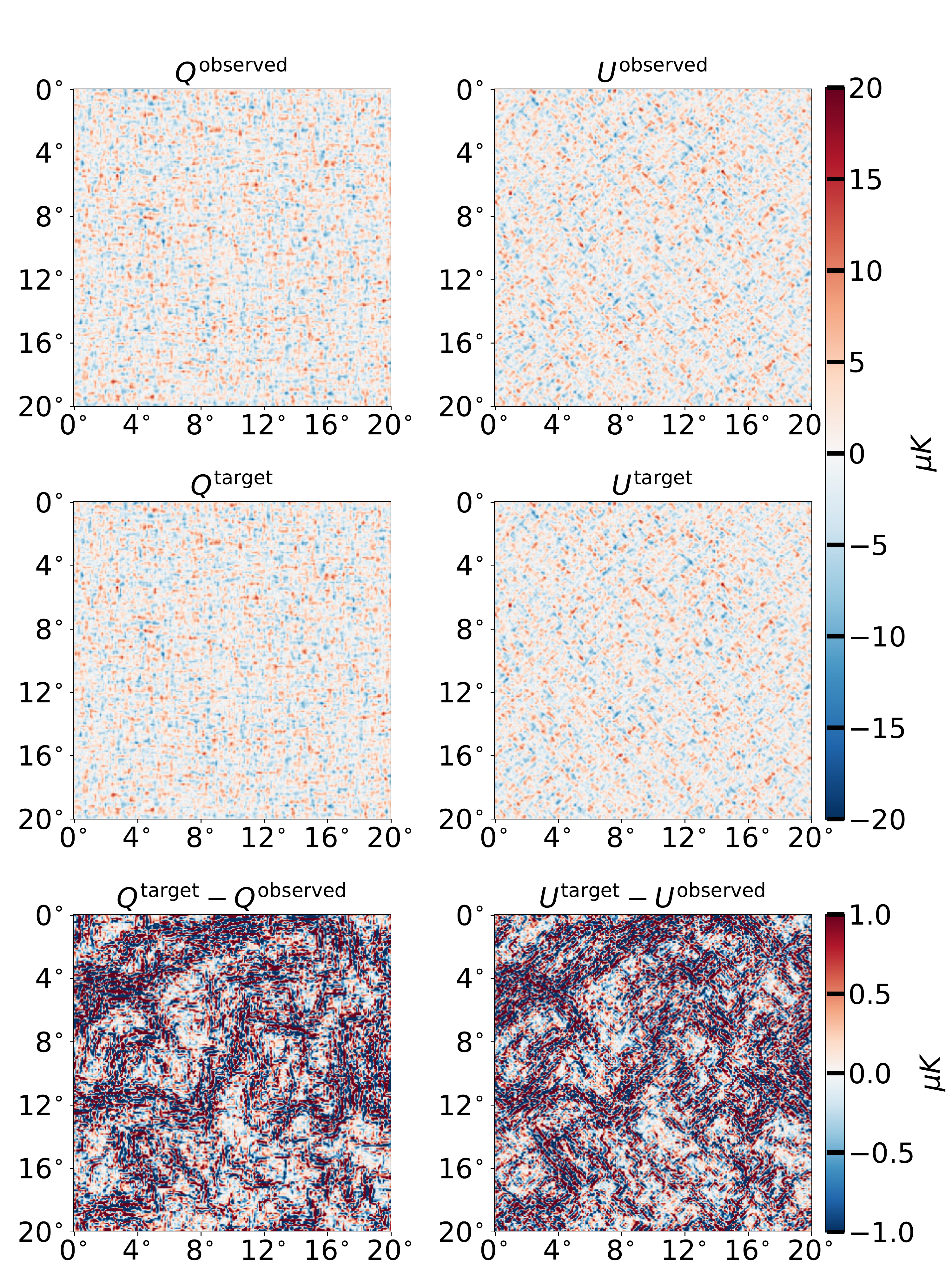}
\end{center}
\vspace{-0.4cm}
\caption{Examples of the observed and target CMB Q/U maps. The residual maps  are shown in the bottom row.}
\label{inputmap}
\end{figure}
\subsection{Data pipeline and network training}
\label{data_pipe}
As a supervised machine learning technique, the CNN method requires a training data set with values that are already known to be truth. The training data set is based on a simulation of a flat sky. We use the \texttt{CAMB}\footnote{\url{https://github.com/cmbant/CAMB}} package to calculate the primordial CMB and lensing potential power spectra \citep{Lewis2001}. Here, we consider a standard $\Lambda$ cold dark matter model with parameters of ($H_0, \Omega_bh^2, \Omega_ch^2, \tau, A_s, n_s$), and their best-fit value and standard deviation can be obtained from the Planck 2015 data \citep{PlanckCollaboration2016}. Additionally, we also take into account the primordial tensor perturbation and set the tensor-to-scalar ratio $r$ to 0.05. In this work, in order to get closer to a realistic manner, the values of cosmological parameters are treated as independent Gaussian random variables, where the mean values and standard deviations are derived from the best-fit values and standard deviation of Planck-2015 result.
\begin{figure*}
\begin{center}
	\includegraphics[width=1\hsize]{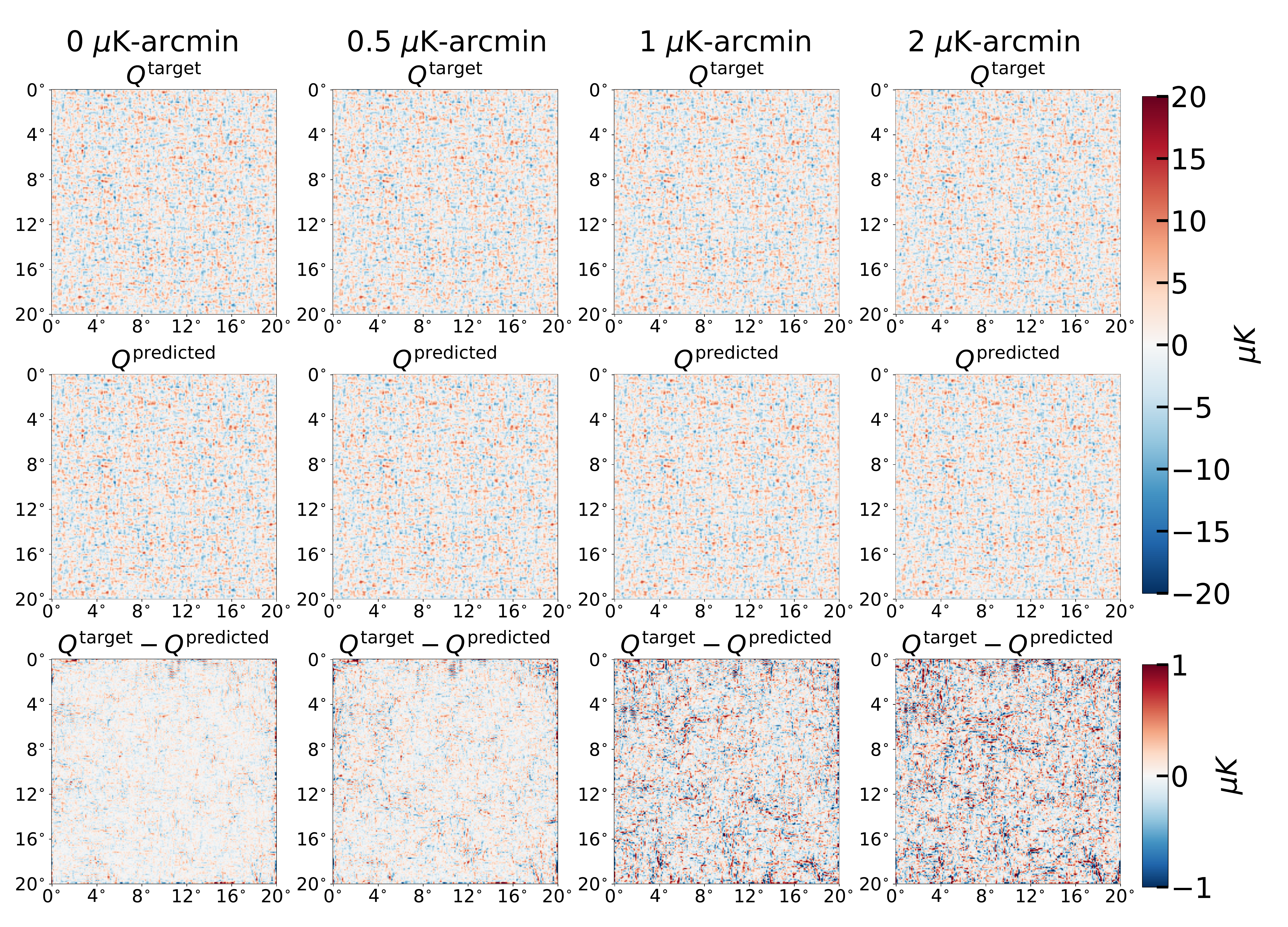}
\end{center}
\vspace{-0.4cm}
\caption{Example of the delensing CMB Q  maps for four noise levels (0, 0.5, 1, 2 $\mu$K-arcmin; left to right). The top row  is the target maps without lensing effect, but with noise and beam information. The middle row is the predicted delening maps by the MIMO-UNet. The bottom row shows the residual maps.}
\label{resultsmap}
\end{figure*}

\begin{table*}
	\centering
	\renewcommand\tabcolsep{15pt}
	\caption{Mean Absolute Error (MAE) between Two Maps over Test Set.\label{table_mae}}
	\begin{tabular}{c|c|c}
		\hline \hline
		\multicolumn{3}{c}{MAE between lensing maps and unlensing maps}\\
		\hline
		-  & MAE for the Q maps & MAE for the  U maps  \\
		-  & ($\mu$K) & ($\mu$K) \\
		\hline
		-  & $1.074\pm0.108$  &  $1.169\pm0.114$ \\
		\hline
		\hline
		\multicolumn{3}{c}{MAE between predicted delensing  maps and target maps}\\
		\hline
		Noise  & MAE for the predicted Q maps & MAE for the predicted  U maps  \\
		($\mu$K-arcmin)  & ($\mu$K) & ($\mu$K) \\
		\hline
		$0$  & $0.133\pm0.008$  &  $0.151\pm0.009$ \\
		$0.5$  & $0.175\pm0.009$  &  $0.194\pm0.009$ \\
		$1$  & $0.286\pm0.015$  &  $0.312\pm0.014$ \\
		$2$  & $0.325\pm0.015$  &  $0.360\pm0.015$ \\
		\hline
		\hline
		\multicolumn{3}{c}{MAE between predicted delensing  maps and target maps for patch size  of $15^{\circ} \times 15^{\circ}$ }\\
		\hline
		Noise  & MAE for the predicted Q maps & MAE for the predicted  U maps  \\
		($\mu$K-arcmin)  & ($\mu$K) & ($\mu$K) \\
		\hline
		$0$  & $0.098\pm0.006$  &  $0.103\pm0.006$ \\
		$1$  & $0.274\pm0.014$  &  $0.286\pm0.013$ \\
		\hline
		\hline
		\multicolumn{3}{c}{MAE between predicted delensing  maps and target maps for patch size  of $10^{\circ} \times 10^{\circ}$ }\\
		\hline
		Noise  & MAE for the predicted Q maps & MAE for the predicted  U maps  \\
		($\mu$K-arcmin)  & ($\mu$K) & ($\mu$K) \\
		\hline
		$0$  & $0.075\pm0.004$  &  $0.068\pm0.004$ \\
		$1$  & $0.331\pm0.020$  &  $0.330\pm0.019$ \\
		\hline
		\hline
		\multicolumn{3}{c}{MAE between reconstructed lensing B map and target lensing B map}\\
		\hline
		Noise  & \multicolumn{2}{c}{MAE for reconstructed lensing B map} \\
		($\mu$K-arcmin)  & \multicolumn{2}{c}{($\mu$K)} \\
		\hline
		$0$  & \multicolumn{2}{c}{$0.028\pm0.001$}  \\
		$1$  & \multicolumn{2}{c}{$0.035\pm0.001$}  \\
		$2$  & \multicolumn{2}{c}{$0.062\pm0.001$}  \\
		\hline \hline
	\end{tabular}
\end{table*}

Based on the these theoretical power spectra from \texttt{CAMB}, we simulate two-dimensional flat sky maps  by using a modified version of \texttt{Orphics}\footnote{\url{https://github.com/msyriac/orphics}} and \texttt{resunet-cmb}\footnote{\url{https://github.com/EEmGuzman/resunet-cmb}}. The simulated flat sky maps cover $20^{\circ} \times 20^{\circ}$  patch of sky with $192 \times 192$ pixels. In total, three different types of maps are generated ($Q^{\rm prim}, U^{\rm prim}, \kappa$). Here, the  `prim' represents the primordial CMB maps and $\kappa$ is the convergence map. These three types of maps can be used to generate observed CMB maps ($Q^{\rm observed}, U^{\rm observed}$), which act as the network's input, and training target maps ($Q^{\rm target}, U^{\rm target}$). In order to obtain the observed maps ($Q^{\rm observed}, U^{\rm observed}$), we firstly lens the primordial maps ($Q^{\rm prim}, U^{\rm prim}$) with the convergence ($\kappa$) map. Then, similar to Refs. \citep{Caldeira2019,Guzman2021},  we smooth these lensed maps with a Gaussian beam size of FWHM= 1 acrmin. Finally, we add a noise map to these beam convolved lensed ($Q$, $U$) maps. Here, we choose four noise levels: $0.0$ $\mu$K-arcmin,  $0.5$ $\mu$K-arcmin, $1$ $\mu$K-arcmin,  and $2$ $\mu$K-arcmin. These instrumental specifications are CMB-S4 like experiment. To obtain training target maps ($Q^{\rm target}, U^{\rm target}$), the primordial maps ($Q^{\rm prim}, U^{\rm prim}$) are directly smoothed with a Gaussian beam and then the noise maps is added. It should be noted that target maps does not implement lensing effect. Here, the noise and beam on these target maps are identical to those on the matching observed maps. Examples of observed and target CMB maps are shown in the Figure \ref{inputmap}.  
For quantitative comparison, we use the mean absolute error (MAE) with the following general formula:
\begin{equation}
	\begin{split}
		\sigma_{\rm MAE} &= \frac{1}{N}\sum_{i}^{N}\left|X_i - Y_i\right|,
	\end{split}
\end{equation}
where $N$ is the number of pixels, $X$ and $Y$ represent the  predicted and target maps. The MAE between observed (lensing) and target CMB Q/U maps (unlensing)  is calculated. As shown in Table \ref{table_mae}, the MAEs are:  $1.074 \pm 0.108\ \mu$K for the Q map and  $1.169 \pm 0.114\ \mu$K for the U map. Due to the lensing effect, we can see that the residual maps retain a lot of information.

For each noise level, 50000 sets of four  independent  maps ($Q^{\rm observed}, U^{\rm observed}, Q^{\rm target}, U^{\rm target}$) are generated and split into training, validation, and test sets with a ratio of 8:1:1. Each noise level is trained using a separate network. In order to train the network, we adopt the Adam \citep{Kingma2014} as the optimizer and initially set the learning rate to 0.001, which gradually decreases to $10^{-6}$ during the iterations. In training, we run 60000 iterations with a batch size of 32. We train our network model on two NVIDIA Quadro GV100 GPUs, and one network model take $\sim$ 27 hours to train. Due to the lensing effect, we can see that the residual maps retain a lot of information.

\section{Delensing CMB polarization maps with the MIMO-Unet model}
\label{delensing_result}

In this section, we present results for delenisng CMB ($Q, U$) maps using the MIMO-Unet model. The delensing CMB maps using the network are denoted by ($Q^{\rm predicted}$, $U^{\rm predicted}$). The observed CMB maps ($Q^{\rm observed}$, $U^{\rm observed}$) are used as input of network, and the desired output is unlensing CMB maps ($Q^{\rm target}$, $U^{\rm target}$) with instrument noise and beam. This means that our network aims to find the mapping between the observed CMB maps and delensing CMB maps, and does not deal with noise and instrumental beam.
\begin{figure*}
\begin{center}
	\includegraphics[width=1\hsize]{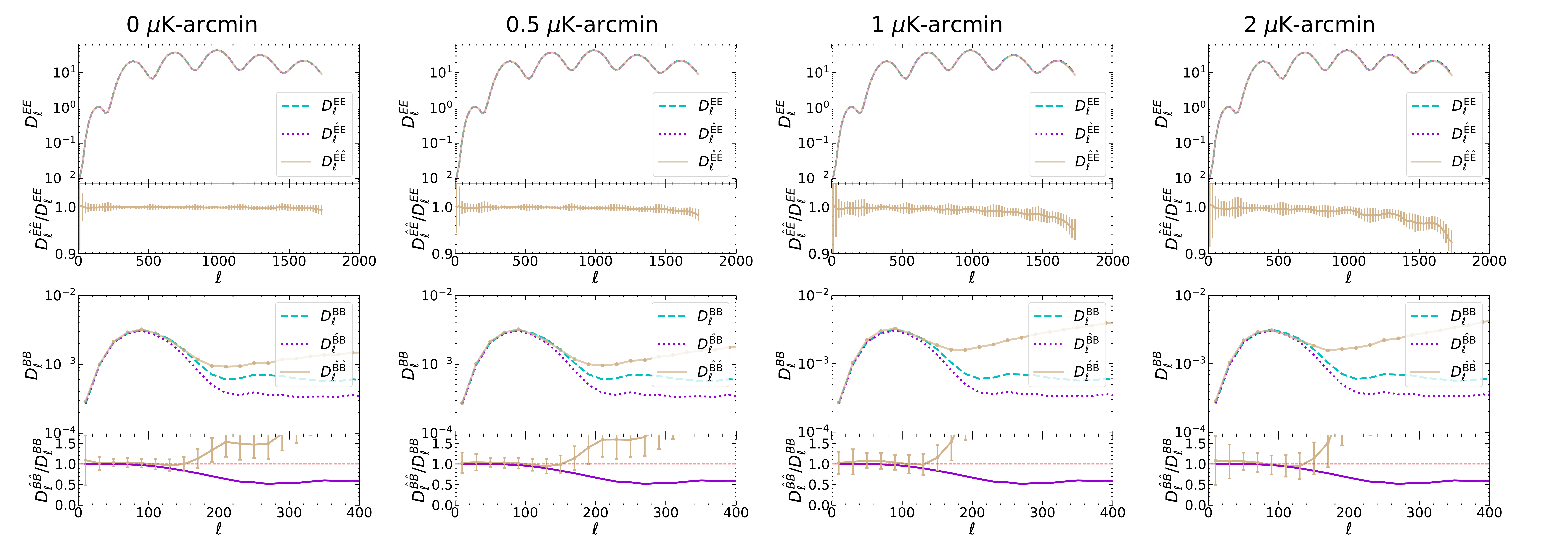}
\end{center}
\vspace{-0.4cm}
\caption{The delensed EE (top panels) and BB (bottom panels) power spectra for four noise levels (0, 0.5, 1, 2 $\mu$K-arcmin; left to right). $\hat{E}$ and $\hat{B}$ represent predicted $E$ and $B$  obtained from our network's output ($Q, U$) maps.  $E$ and $B$ represent the target $E$ and $B$. $D_{\ell}^{\rm \hat{E}E}$ (or $D_{\ell}^{\rm \hat{B}B}$) represents the cross-spectrum of predicted $\hat{E}$ (or $\hat{B}$) and target $E$ (or $B$).  The length of each $\ell$ bin is set to be 20 here. Here,  $D_{\ell} = \ell (\ell+1)C_{\ell}/2\pi$. We adopt the standard deviation on the test set as the error of the estimate.}
\label{denoise_power}
\end{figure*}
\subsection{Delensing results}
\label{Delensing_results}
Our delensing Q maps (U maps have similar results) on the test set are shown in Figure \ref{resultsmap} for the experiments of four levels of noise. From visual inspection, it is hard to distinguish between the predicted Q maps by network and target maps. The bottom panels show the difference between the predicted maps and targets. In the case of low noise levels ($\leq 1\mu$K-arcmin), residual maps have minimal information left, implying that the MIMO-UNet can delens the Q maps very well. The residual maps have more information at 2 $\mu$K-arcmin noise levels than at low noise levels, yet the MIMO-UNet still has good delening performance when compared to the residual maps in Figure  \ref{inputmap}. Furthermore, we can see that all amplitudes of residual maps grow visibly as noise levels increase, implying that delensing performance decreases as noise level increases. In order to evaluate the performance of our network quantitatively, we calculate the MAE between predicted delensing and target Q maps shown in Figure  \ref{resultsmap}: $0.134 \ \mu$K, $0.178\ \mu$K,  $0.282\ \mu$K,  $0.322\ \mu$K for 0 $\mu$K-arcmin, 0.5 $\mu$K-arcmin, 1 $\mu$K-arcmin, and 2 $\mu$K-arcmin noise levels, respectively. In addition,  as shown in Table \ref{table_mae}, the average MAEs over the testing sets for delensing Q maps are  $0.133 \pm 0.008\ \mu$K, $0.175 \pm 0.009\ \mu$K, $0.286 \pm 0.015\ \mu$K, and $0.325 \pm 0.015\ \mu$K for 0 $\mu$K-arcmin, 0.5 $\mu$K-arcmin, 1 $\mu$K-arcmin, and 2 $\mu$K-arcmin noise levels, respectively. The MAE for delensing U maps  are also listed by Table \ref{table_mae}. These quantitative comparisons also support our conclusions.

To make these observations more precise, we compute the power spectrum of delensing CMB maps and compare them to the primordial EE and BB spectra. In our method, the output CMB Q/U maps from the MIMO-UNet still contain the instrument noise. According to \cite{Wang2022} and \cite{Yan2023}, it is still very challenging to directly remove noise effects at the map level using CNN methods. Therefore, in this network we leave the noise in the output maps and try to suppress the noise effect at the power spectrum level. Inspired by \citep{Krachmalnicoff2022}, we divide the whole dataset into two "half-split" (HS) maps which share the same signal but have uncorrelated noise, and then calculate the cross-correlation power spectrum of these two HS maps. Due to their uncorrelated noise, the noise effect almost disappears in their cross-correlation power spectrum, but the CMB signal remains. It should be noted that the noise of HS maps is enhanced by a factor of $\sqrt{2}$ relative to the primordial noise level. 

We use \texttt{NaMaster}\footnote{\url{https://github.com/LSSTDESC/NaMaster}} \citep{Alonso2019} to calculate the E- and B-mode power spectra from the predicted Q and U maps by the MIMO-UNet. Figure \ref{denoise_power} shows the delensed EE and BB spectra after denoised step for four noise levels, and these power spectra are averaged over the test set.  The reconstructed $EE$ spectra ($D_{\ell}^{\hat{E}\hat{E}}$) are very consistent with primordial EE spectra ($D_{\ell}^{EE}$) across the entire range of scales we consider. Adding noise to the input maps does not degrade the recovery of EE power spectrum significantly at all the noise levels, which is consistent with results of \citep{Caldeira2019,Guzman2021}. The cross-correlation spectra ($D_{\ell}^{\hat{E}E}$)  of reconstructed $\hat{E}$ and true $E$ is also calculated. We can see that $D_{\ell}^{\hat{E}E}$  are very consistent with $D_{\ell}^{\hat{E}\hat{E}}$, which could imply that our reconstructed E-mode signal is the true signal. 

The recovery of BB spectra ($D_{\ell}^{\hat{B}\hat{B}}$) gets systematically worse as $\ell$ increases. Specifically, the recovery of the B-mode power spectrum is greater than about 98\% of the primordial B-mode at the angular scale of $\ell<150$, but it gradually deviates from the primordial BB ($D_{\ell}^{BB}$) as $\ell$ increases at the scales of $\ell>150$. In particular, the recovered B-mode power spectrum significantly deviates from  the primordial BB at the scales of $\ell>200$. In fact, the lensing B-mode is at least one order of magnitude higher than the primordial BB at the scales of $\ell>200$,  implying that recovering the primordial B-mode is extremely challenging. Therefore, we think that  the recovered B-mode power spectrum by our network should be dominant by lensing B-mode at the scales of $\ell>200$ because our network does not completely remove the lensing effect at these scales. A more detailed analysis will be presented in section \ref{null_test_lensi}. The cross-correlation spectra ($D_{\ell}^{\hat{B}B}$)  of reconstructed $\hat{B}$ and true $B$ is also calculated. We can see that $D_{\ell}^{\hat{B}B}$  are consistent with $D_{\ell}^{\hat{B}\hat{B}}$ at the scales of $\ell<140$, but $D_{\ell}^{\hat{B}B}$ gradually deviates from the $D_{\ell}^{\hat{B}\hat{B}}$ at the scales of $\ell>140$. In particular, at the scales of $\ell>200$, $D_{\ell}^{\hat{B}B}$ is about 60-50\% of $D_{\ell}^{BB}$. These could suggest that our recovered BB spectra ($D_{\ell}^{\hat{B}\hat{B}}$) are true primordial B-mode signals at the scales of $\ell>140$, but the primordial B-mode signal contained in $D_{\ell}^{\hat{B}\hat{B}}$ will gradually lose as $\ell$ increases.

\subsection{Null test}
\label{null_test_lensi}
Now, we need to check that the network encodes a sensible mapping of the input observed maps to delensing maps, i.e., the reconstructed BB spectrum by the network model is due to the presence of the real tensor BB rather than an artifact of the network and it is the real tensor BB rather than a reduced lensing BB. In addition, we also wish to understand why the reconstructed tensor BB deviates from the primordial BB spectrum at the scales of $\ell>150$. Based on these considerations, two null tests are performed. 

\begin{figure}
\begin{center}
	\includegraphics[width=1\hsize]{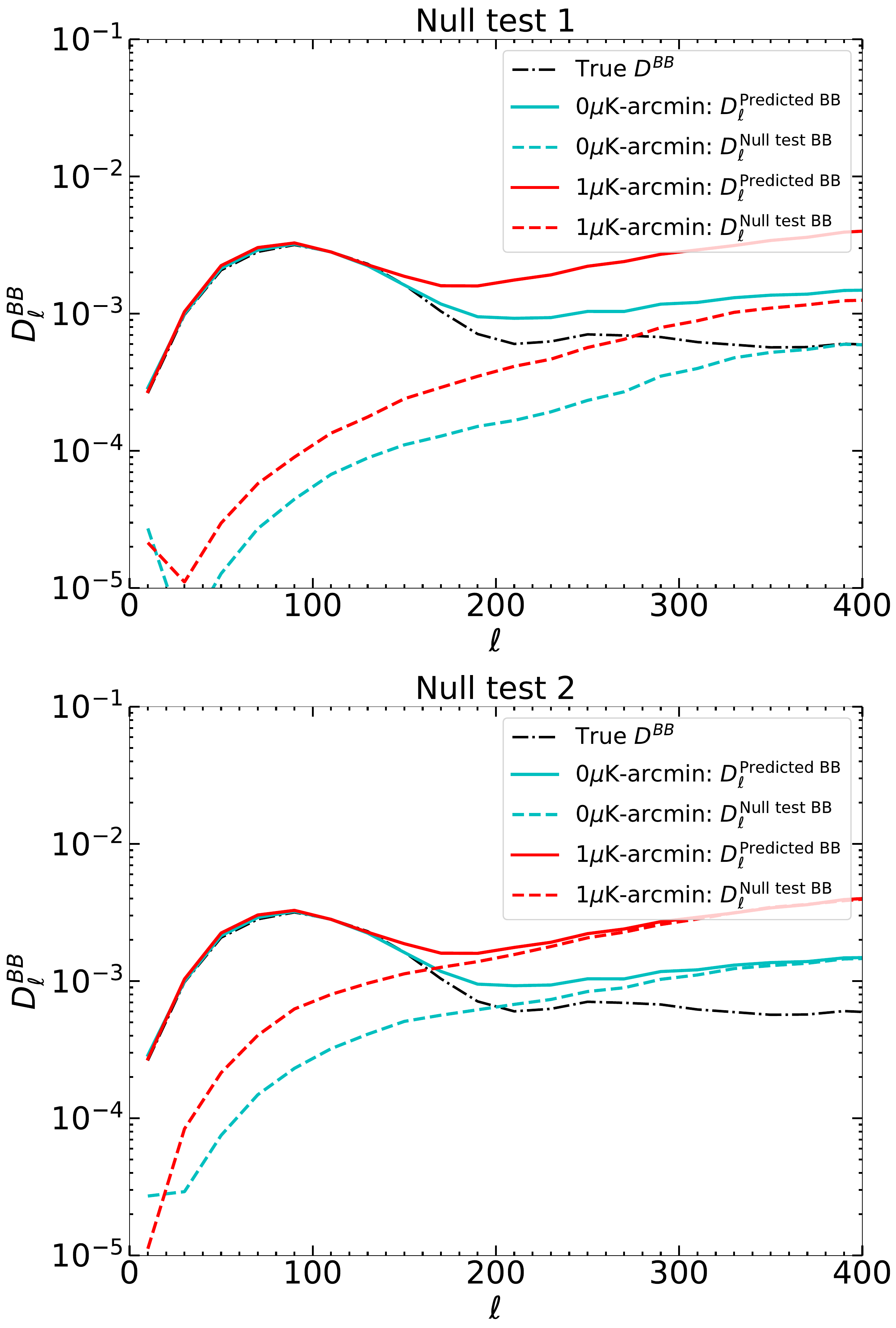}
\end{center}
\vspace{-0.4cm}
\caption{Null test for experiments of 0 and 1 $\mu$K-arcmin noise levels. Top panel: null test 1 represents that we feed unlensed versions of ($Q$,$U$) maps with $r=0$ to MIMO-Unet model trained on the observed ($Q^{\rm observed}, U^{\rm observed}$) maps. Bottom panel: null test 2 represents that we feed lensed versions of ($Q$, $U$) maps with $r=0$ to the network model trained on the observed ($Q^{\rm observed}, U^{\rm observed}$) maps. $D_{\ell}^{\rm Predicted\ BB}$ and $D_{\ell}^{\rm Null\ test\ BB}$ are power spectrum of normal network output shown in Figure \ref{denoise_power} and power spectrum from null test output, respectively.}
\label{null_test}
\end{figure}

For the first null test (null test 1), we feed the unlensed versions of ($Q$,$U$) maps with no tensor-to-scalar ratio (i.e., $r=0$) into the network model trained on the observed ($Q^{\rm observed}, U^{\rm observed}$) maps. Here, the noise and beam on these unlensed versions maps are the same as those on observed ($Q^{\rm observed}, U^{\rm observed}$) maps. We calculate the power spectrum of the output ($Q^{\rm pred}, U^{\rm pred}$) from this null test. The upper panel of Figure \ref{null_test} shows the predicted power spectrum from this null test for noiseless and 1 $\mu$K-arcmin noise experiments. The predicted BB spectrum for 1 $\mu$K-arcmin noise is denoised by cross-correlation of two HS maps. We can see that the predicted BB spectra from the null test look like the noise signals, and predicted $D_{\ell}^{BB}$ for 1 $\mu$K-arcmin noise level is higher than that of noiseless experiment. These imply that the output ($Q^{\rm pred}, U^{\rm pred}$) could be the random noise when we input the unlensed (Q,U) maps with  $r=0$. Furthermore, the amplitude of this output noise depends on the input noise level. 
\begin{figure*}
\begin{center}
	\includegraphics[width=0.9\hsize]{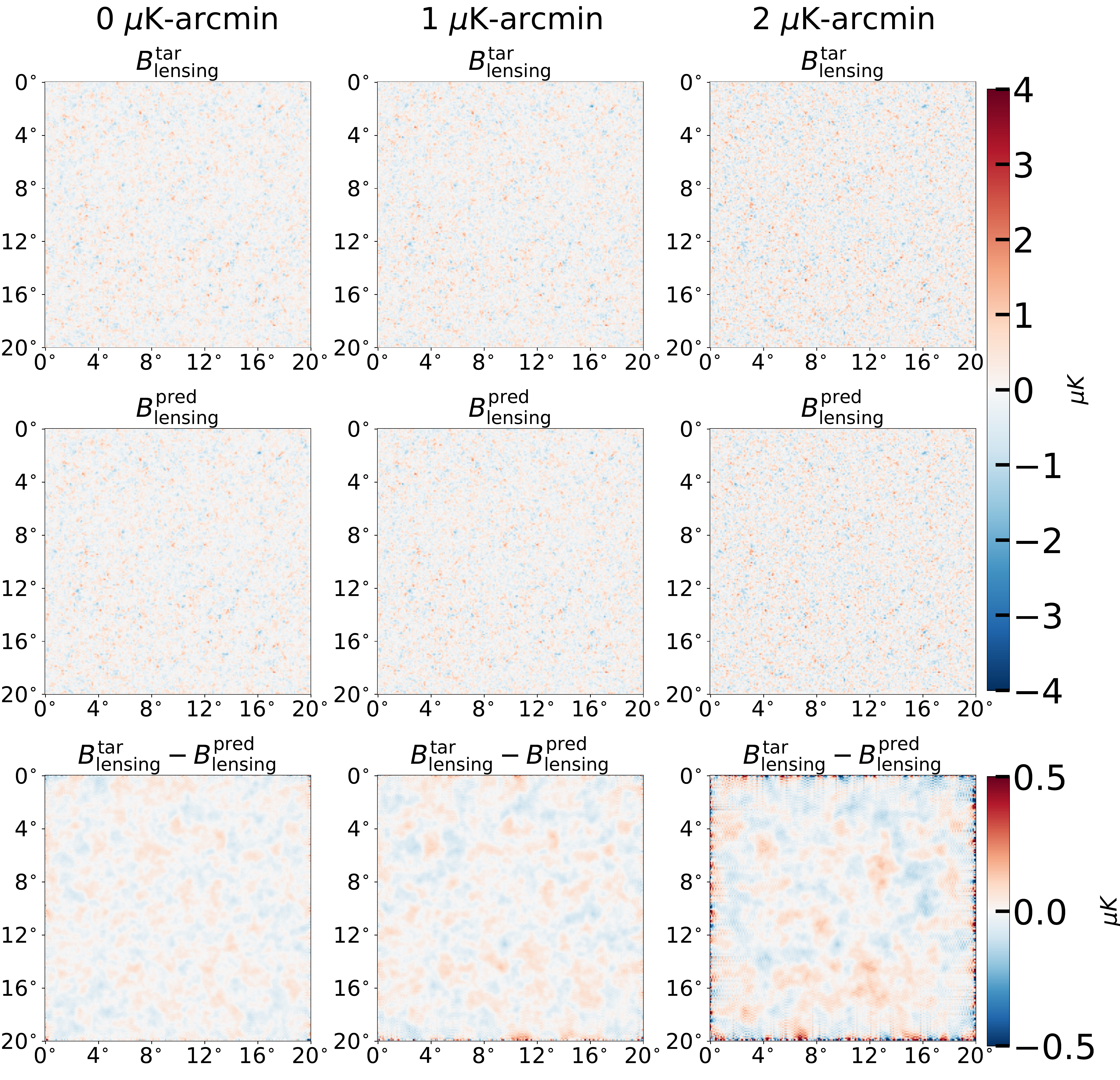}
\end{center}
\vspace{-0.4cm}
\caption{Example of reconstructing lensing B maps for three noise levels (0, 1, 2 $\mu$K-arcmin; left to right). The top row  is the target lensing B maps ($B_{\rm lensing}^{\rm tar}$, that is, observed lensing B map $B^{\rm obs\ lensing}$). The middle row is the predicted lensing B maps ($B_{\rm pred}^{\rm tar}$) by the MIMO-UNet. The bottom row shows the related residual maps.}
\label{results_B_map}
\end{figure*}

Next, we perform the second null test (null test 2) and feed lensed versions of ($Q$, $U$) maps with $r=0$ to the network model trained on the observed ($Q^{\rm observed}, U^{\rm observed}$) maps. The noise and beam on these lensed versions maps are the same as those on observed ($Q^{\rm observed}, U^{\rm observed}$) maps. The bottom panel of Figure \ref{null_test} shows that the power spectra of the output ($Q^{\rm pred}, U^{\rm pred}$) from this null test for noiseless and 1 $\mu$K-arcmin noise experiments. We can see that the predicted BB spectra from null test 2 still look like the noise signals, which suggests that the reconstructed BB spectrum by the network model is from the real tensor BB. In addition, since the input map contains lensing effect, BB spectra from null test 2 are higher than those null test 1. We can understand this as: our network has not completely removed the lensing effect, that is, there is still a small amount of lensing on the output map of our network, which can also be observed on the residual maps of Figure \ref{resultsmap}. These small amounts of lensing on the output map will  contribute to the BB spectrum ($D_{\ell}^{BB,{\rm reduced\ lensing}}$). As a result, the predicted BB spectrum ($D_{\ell}^{BB,{\rm predicted}}$) from null test 2 can be modeled as 
\begin{align}
	D_{\ell}^{BB,{\rm predicted}} = D_{\ell}^{BB,{\rm reduced\ lensing}} + N_{\ell},
\end{align}
where $N_{\ell}$ is the noise spectrum. We also plot the delensing BB spectrum by network from observed ($Q^{\rm observed}, U^{\rm observed}$) maps in the bottom panel of Figure \ref{null_test}. We see that predicted BB spectra by our network are consistent with the results of null test 2 at the angular scales of $\ell>300$ for the noiseless experiment and $\ell>250$ for 1 $\mu$K-arcmin noise experiment. These could suggest that the reconstructed tensor BB deviates from the primordial BB spectrum at the scales of $\ell>150$ as shown in Figure \ref{denoise_power} due to the contributions of $D_{\ell}^{BB,{\rm reduced\ lensing}}$ and $N_{\ell}$.

\section{Reconstruct lensing B map}
\label{Reconstruct_lensing}
\subsection{Reconstructed results}
\label{Reconstruct_lensing_results}

In section \ref{Delensing_results}, we use the CNN to delens Q and U maps. Inspired by \cite{Manzotti2017}, we now try to reconstruct the lensing B mode using the CNN, and we delens observed B-mode power spectrum by subtracting this reconstructed lensing B-mode power spectrum. 
\begin{figure*}
\begin{center}
	\includegraphics[width=1\hsize]{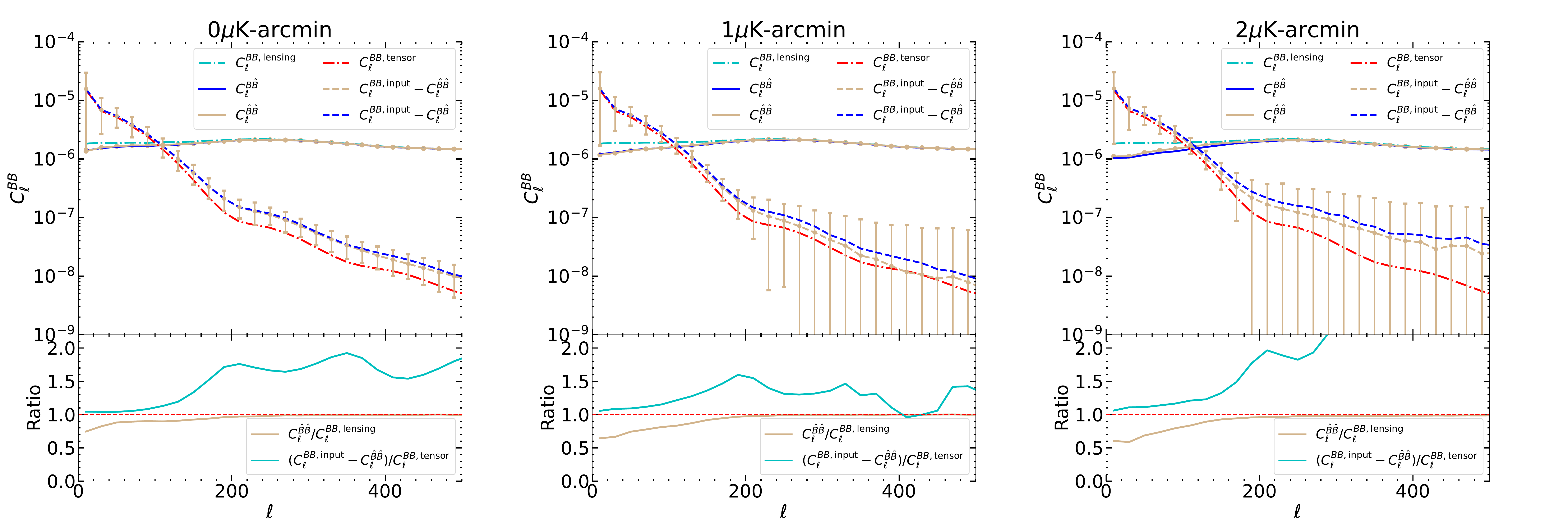}
\end{center}
\vspace{-0.4cm}
\caption{The recovered BB power spectra for experiments of three noise levels (0, 1, 2 $\mu$K-arcmin; left to right). $B$ and $\hat{B}$ represent the true lensing B map and predicted lensing B map by the network. $C_{\ell}^{\rm BB,lensing}$ and $C_{\ell}^{\rm BB,tensor}$ are true lensing BB spectrum and primordial tensor BB spectrum. $C_{\ell}^{B\hat{B}}$ is cross-correlation between $B$ and $\hat{B}$. The $C_{\ell}^{\hat{B}\hat{B}}$ is predicted lensing BB spectrum by network. $C_{\ell}^{\rm BB,input}$ is network input BB spectrum from observed CMB $Q$ and $U$ maps. The power spectra are averaged over the test set, and we adopt the standard deviation on the test set as the error of the estimate of the tensor BB spectrum.}
\label{results_BB_denoise_power}
\end{figure*}
Similar to the simulation process in Section \ref{data_pipe}, we simulate five different types of maps in total, ($Q^{\rm prim}$, $U^{\rm prim}$, $Q^{\rm prim, r=0}$,  $U^{\rm prim, r=0}$,  $\kappa$). Here, the  `prim' and `prim,r=0' represent the primordial CMB maps with $r=0.05$ and primordial CMB maps with $r=0$. $\kappa$ is the convergence map. Based on these five types of maps, the observed maps ($Q^{\rm observed}$, $U^{\rm observed}$, $Q^{\rm observed,r=0}$, $U^{\rm observed,r=0}$) and training target maps of the network can be generated. Here, the `observed' and `observed,r=0' represent the observed CMB maps with r=0.05 and observed CMB maps with r=0. To obtain observed maps, we firstly lens the primordial maps (($Q^{\rm prim}, U^{\rm prim}$) or ($Q^{\rm prim, r=0}$,  $U^{\rm prim, r=0}$)) with the convergence ($\kappa$) map. Then, we smooth these lensed maps with a Gaussian beam size of FWHM= 1 acrmin. Finally, we add a noise map to these beam convolved lensed ($Q$, $U$) maps. Here, we choose three levels of noise: $0.0$ $\mu$K-arcmin, $1$ $\mu$K-arcmin,  and $2$ $\mu$K-arcmin. Note that the observed maps with $r=0.05$ ($Q^{\rm observed}$, $U^{\rm observed}$) take as input of network. In this section, the desired output of the network is observed lensing B map ($B^{\rm obs\ lensing}$). Here, $B^{\rm obs\ lensing}$ map is generated using observed CMB maps with $r=0$ ($Q^{\rm observed,r=0}$, $U^{\rm observed,r=0}$). This means that the network is designed to reconstruct lensing B map from observed ($Q^{\rm observed}$, $U^{\rm observed}$) maps. For each noise level, 50000 sets of three independent  maps ($Q^{\rm observed}$, $U^{\rm observed}$,  $B^{\rm obs\ lensing}$) are generated and split into training, validation, and test sets with a ratio of 8:1:1. Each noise level is trained using a separate network.

Figure \ref{results_B_map} shows the reconstructed lensing B maps for three noise levels (0, 1, 2 $\mu$K-arcmin) on the test set. From visual inspection, we see that there is little information left on the residual map, implying that the network can reconstruct lensing B map fairly well. However, as the levels of noise increasing, all the amplitudes of residual maps increase. This implies that the performance of reconstructing lensing B map by network depends on the noise level of the observed map. The MAEs between predicted lensing and target lensing B maps as shown in Figure \ref{results_B_map} is calculated for the experiments of three levels of noise: $0.029\ \mu$K, $0.036\ \mu$K, and $0.061 \mu$K for 0 $\mu$K-arcmin, 1 $\mu$K-arcmin,  2 $\mu$K-arcmin noise levels, respectively. In addition, as shown in Table \ref{table_mae}, the average MAEs over the testing sets are  $0.028 \pm 0.001\ \mu$K, $0.035 \pm 0.001\ \mu$K, and $0.062 \pm 0.001\ \mu$K for 0 $\mu$K-arcmin, 1 $\mu$K-arcmin,  2 $\mu$K-arcmin noise levels, respectively. These quantitative comparisons also support our conclusions.

Next, we present the power spectrum of the reconstructed lensing B map at each noise level. Figure \ref{results_BB_denoise_power} shows the power spectra of reconstructed lensing B map for three noise levels: noiseless, 1 $\mu$K-arcmin, 2 $\mu$K-arcmin. These power spectra are the mean result over the test set. Note that the reconstructed power spectra by the network are denoised by the cross-correlation of HS maps. At the angular scales of $\ell>200$, the reconstructed lensing BB power spectra ($C^{\hat{B}\hat{B}}_{\ell}$) by our network are quite consistent with the truth lensing BB spectra ($C^{BB,{\rm lensing}}_{\ell}$). The recoveries of lensing BB power spectra are greater than about 99\% and adding noise to input maps does not degrade lensing-B mode recovery significantly from the noiseless case. However, the reconstructed lensing BB ($C^{\hat{B}\hat{B}}_{\ell}$) slightly deviates from the truth lensing BB at the angular scales of $\ell<200$, which could be affected by tensor B-mode. The recovery of the lenging BB at the angular scales of $\ell<200$ visibly degrades when noise is added to the input maps. This implies that the noise has a negative impact on lensing BB spectra reconstruction, which is consistent with the obtained results of the reconstructed lensing B map in Figure \ref{results_B_map}.

We also plot the cross-correlation spectrum ($C^{B\hat{B}}_{\ell}$) between the reconstructed lensing B map and true lensing B map in Figure \ref{results_BB_denoise_power}. We see that reconstructed BB spectra are consistent with $C^{B\hat{B}}_{\ell}$ for all noise levels, which implies that our reconstructed lensing B map is a true lensing B mode signal. 

Next, we delens our observed BB power spectrum by subtracting the reconstructed lensing BB spectrum. This `spectrum difference' is defined as 
\begin{align}
	\label{estimate}
	\Delta C^{\hat{B}\hat{B}}_{\ell} \equiv &  C^{BB, {\rm obs}}_{\ell} - C^{\hat{B}\hat{B}}_{\ell}.
\end{align}
We use this $\Delta C^{\hat{B}\hat{B}}_{\ell}$ as the estimate of the tensor BB spectrum. We also plot the $\Delta C^{\hat{B}\hat{B}}_{\ell}$ (yellow dotted line) for four noise levels in Figure \ref{results_BB_denoise_power}. We see that $\Delta C^{\hat{B}\hat{B}}_{\ell}$ are basically consistent with tensor BB spectrum ($C^{BB,{\rm tensor}}$) at the angular scales of $\ell<180$, and adding noise to input maps does not degrade tensor-B mode recovery significantly.  However, $\Delta C^{BB}_{\ell}$  deviates from the truth tensor BB at the angular scales of $\ell>180$. More specifically, the B-mode power spectrum recovery at noise levels of 0, 1, 2 $\mu$K-arcmin is greater than about 98\% at the scales of $\ell<120$. Tensor BB power spectrum recovery is around 71\% at $\ell=160$. We also plot the difference between the observed BB power spectrum and cross-correlation spectrum in Figure \ref{results_BB_denoise_power}, which is defined as 
\begin{align}
	\Delta C^{B\hat{B}}_{\ell} \equiv &  C^{BB, {\rm obs}}_{\ell} - C^{B\hat{B}}_{\ell}.
\end{align}
We see that $\Delta C^{B\hat{B}}_{\ell}$ are consistent with $\Delta C^{\hat{B}\hat{B}}_{\ell}$ at the angular scales of $\ell<200$ for all noise levels.

We note that the errors of the recovered tensor B-mode power spectrum increase sharply at the angular scales of $\ell>220$ for the experiment of 1 $\mu$K-arcmin noise level and $\ell>190$ for the experiment of 2 $\mu$K-arcmin noise level, while error for the noiseless experiment does not increase significantly with the increase of $\ell$. This implies that the increase in error could be due to the presence of noise. In fact, the reconstructed map by network model contains the instrumental noise, and we always hope that we can use the cross-correlation technique to suppress the instrumental noise effect on the power spectrum. However, it is difficult to suppress noise on small scales because the tensor B-mode signal is very weak. This could lead to an increase in the error of  the recovered tensor B-mode power spectrum at the small scales. Specifically, the reconstructed $\hat{B}$ map can be written as
\begin{align}
	\hat{B}_{\ell} = B_{\ell} +N_{l}, 
\end{align}
where $N_{l}$ and $B_{\ell}$ are noise and true B-mode field. The cross-correlation of the two reconstructed HS maps, $\hat{B}^{1}$ and $\hat{B}^{2}$, can be calculated as 
\begin{align}
	\langle \hat{B}_{\ell}^{1} \hat{B}_{\ell}^{2} \rangle = &\langle B_{\ell}B_{\ell} \rangle + \langle B_{\ell} N^1_{\ell} \rangle  + \langle B_{\ell} N^2_{\ell} \rangle  + \langle N^1_{\ell} N^2_{\ell} \rangle.
\end{align}
Here, $\langle \hat{B}_{\ell}^{1} \hat{B}_{\ell}^{2} \rangle$ is an unbiased estimate of $\langle B_{\ell} B_{\ell} \rangle$. We can calculate the variance of $\langle \hat{B}_{\ell}^{1} \hat{B}_{\ell}^{2} \rangle$  as 

\begin{align}
	\langle (\hat{B}_{\ell}^{1} \hat{B}_{\ell}^{2} -B_{\ell}B_{\ell})^2 \rangle = & \langle \hat{B}_{\ell}^{1} \hat{B}_{\ell}^{2}\hat{B}_{\ell}^{1}\hat{B}_{\ell}^{2} \rangle
	+ (C_{\ell}^{BB})^2\\ &- 2 \langle \hat{B}_{\ell}^{1} \hat{B}_{\ell}^{2} \rangle C_{\ell}^{BB}\\
	= &
	C_{\ell}^{BB} (N_{\ell}^{1,1} +N_{\ell}^{2,2}) +N_{\ell}^{1,1}N_{\ell}^{2,2},
\end{align}
where $N_{\ell}^{1,1}$ and $N_{\ell}^{2,2}$ are auto-correlation spectra of noise $N^1$ and $N^2$.  $N_{\ell}^{1,1}N_{\ell}^{2,2}$ is the main source of variance of cross-correlation at about scale of $\ell>200$ for noise level of 1 $\mu$K-arcmin and $\ell>160$ for noise level of 2 $\mu$K-arcmin. Therefore, the errors of the recovered tensor B-mode power spectrum after denoised step using cross-correlation of HS maps increase sharply at small scale.

\begin{figure}
\begin{center}
	\includegraphics[width=1\hsize]{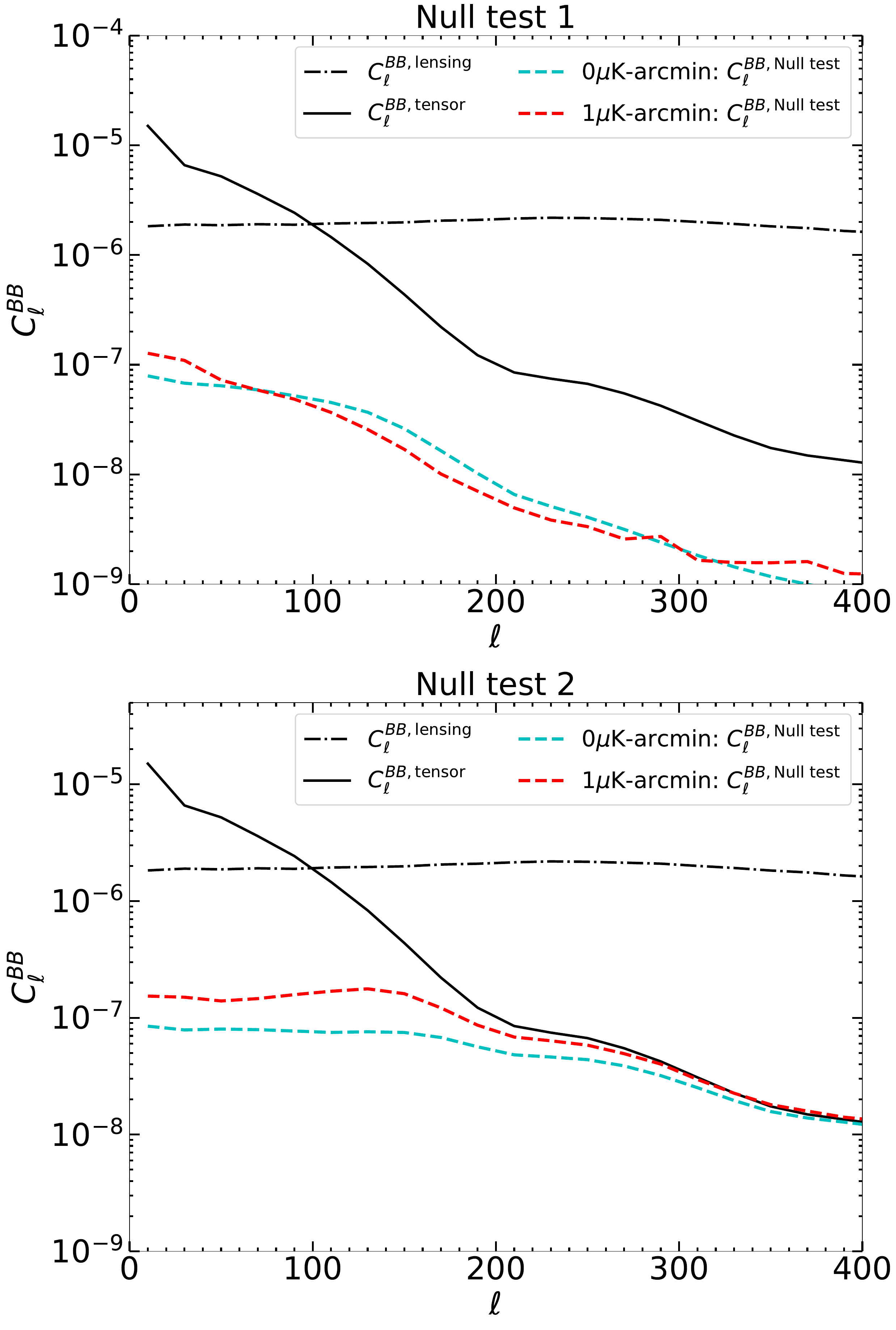}
\end{center}
\vspace{-0.4cm}
\caption{Null test for experiments of 0 and 1 $\mu$K-arcmin noise levels. Top panel: null test 1 represents that we feed unlensed versions of ($Q$,$U$) maps with $r=0$ to MIMO-Unet model trained on the observed ($Q^{\rm observed}, U^{\rm observed}$) maps. Bottom panel: null test 2 represents that we feed unlensed versions of ($Q, U$) maps with $r=0.05$ to the network model trained on the observed ($Q^{\rm observed}, U^{\rm observed}$) maps. $C_{\ell}^{\rm BB,lensing}$ and $C_{\ell}^{\rm BB,tensor}$ are true lensing BB spectrum and primordial tensor BB spectrum.  $C_{\ell}^{\rm Null\ test\ BB}$ is the power spectrum from null test output.}
\label{null_test_lensing}
\end{figure}

\subsection{Null test}
\label{nulltest}
Similar to section \ref{null_test_lensi}, to verify that the network encodes a sensible mapping of the input observed map to lensing B map, we apply two null tests.

For the first null test (null test 1), we feed unlensed versions of ($Q, U$) maps with no tensor-to-scalar ratio (i.e., $r=0$) to the network model trained on the observed ($Q^{\rm observed}$, $U^{\rm observed}$) maps. Here, the noise and beam on these unlensed versions maps are the same as those on observed ($Q^{\rm observed}, U^{\rm observed}$) maps. We consider two experiments with noiseless and 1 $\mu$K-arcmin noise. The power spectrum  of the output from this null test is shown in the upper panel of Figure \ref{null_test_lensing}.  We can see that our reconstructed lensing BB power spectra  from the null test have a small signal and  look like the noise signals. In addition, the reconstructed lensing BB from the null test  is one to three orders of magnitude smaller than the true lensing power spectrum. This implies that our reconstructed lensing B map is true lensing B signal, rather than producing artifacts from the training process. In addition, our reconstructed lensing BB power spectra from the null tests (i.e., noise signal from the trained network) are one to two orders of magnitude smaller than the tensor BB spectrum, which means that this noise signal from the network has little effect on the reconstructed tensor BB power spectrum.

\begin{figure*}
\begin{center}
	\includegraphics[width=1\hsize]{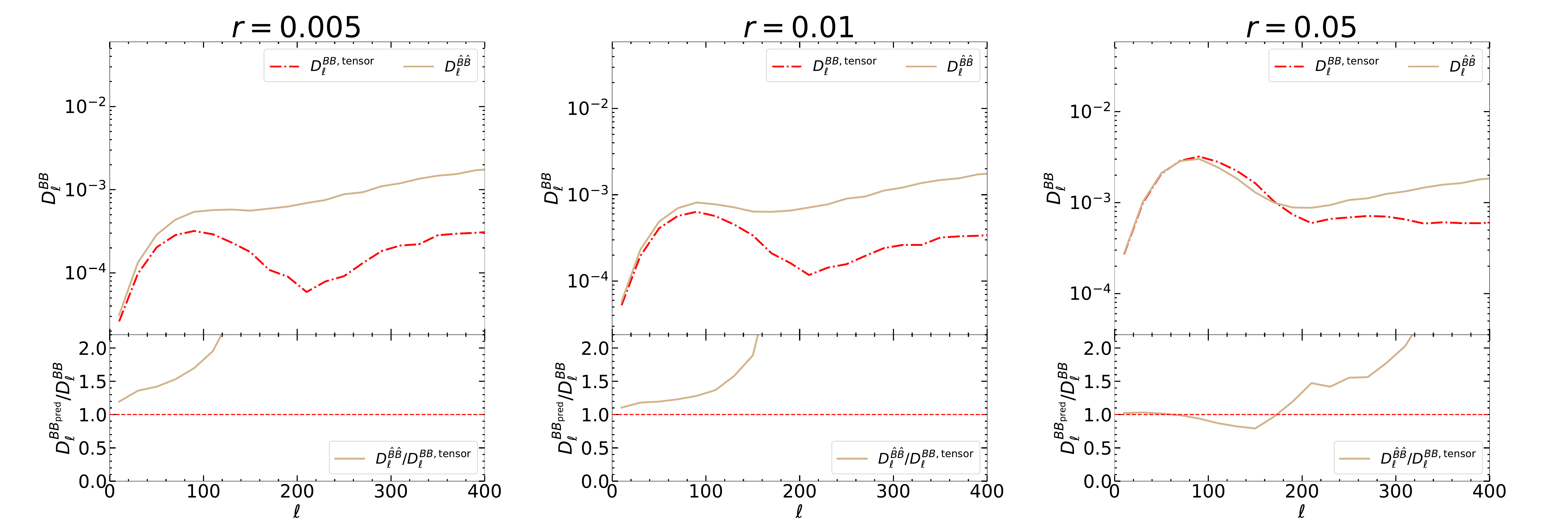}
\end{center}
\vspace{-0.4cm}
\caption{The delensed BB power spectra for experiments of three tensor-to-scalar ratio levels ($r=0.005,0.01,0.05$; left to right). Predicted power spectra by the network are denoised by the cross-correlation of HS maps and averaged over the test set. $D_{\ell}^{\rm BB,tensor}$ and $D_{\ell}^{\rm \hat{B}\hat{B}}$ are primordial tensor  BB spectrum and  predicted BB spectrum by network. We set noise level to 1 $\mu$K-arcmin.}
\label{results_multi_QU_denoise_power}
\end{figure*}
For the second null test, we feed unlensed versions of ($Q, U$) maps with $r=0.05$ to the network model trained on the observed ($Q^{\rm observed}$, $U^{\rm observed}$) maps. Here, the noise and beam on these unlensed versions maps are the same as those on observed ($Q^{\rm observed}, U^{\rm observed}$) maps. The power spectrum of output maps from this null test for experiments of noiseless and 1 $\mu$K-arcmin noise are shown in the bottom panel of Figure \ref{null_test_lensing}. The predicted lensing BB power spectra from this null test also have a small signal similar to noise and are one to two orders of magnitude smaller than the true lensing power spectrum. In addition, this noise signal is higher than noise signal from null test 1 because the tensor-to-scalar ratio is added to the test set of the second null test. We also compare this noise signal from null test 2 output to the tensor BB spectrum as shown in the bottom panel of Figure \ref{null_test_lensing}.  This noise signal is one to two orders of magnitude smaller than the tensor BB spectrum at the angular scales of $\ell<100$. The difference between the two signals gradually decreases at the angular scales of $100<\ell<300$, and these two signals are almost identical at the angular scales of $\ell>300$. These imply that the tensor B signal will have a negative impact on the reconstruction of lensing B map. Our network cannot recognize the tensor BB signal from the observed B mode signal (lensing B model plus tensor B mode) at the angular scales of $\ell>200$ because the tensor BB is at least one order of magnitude smaller than lensing BB. This suggests that our method cannot accurately recover the tensor BB spectrum using E.q. \ref{estimate} at the angular scales of $\ell>200$. This conclusion has been demonstrated by our above results as shown in Figure \ref{results_BB_denoise_power}, where our recovered tensor BB spectrum using E.q. \ref{estimate} is higher than true tensor BB spectrum at the small angular scales of $\ell>150$.

\section{DISCUSSIONS}
\label{DISCUSSIONS}
\subsection{Variability of the tensor-to-scalar ratio}
In section \ref{delensing_result} and \ref{Reconstruct_lensing}, we presented the results of delening CMB polarization maps and reconstructing lensing B map based on the training set with a fixed tensor-to-scalar ratio of 0.05. However, in real observation, we do not know what is its tensor-to-scalar ratio. To enable the network to handle unknown tensor-to-scalar ratio data,  we now consider the variation of the tensor-to-scalar ratio ($r$) on the training set. Similar to the simulation process described in Section \ref{data_pipe}, we here set the tensor-to-scalar ratio to vary in the range of $[0.0001, 0.1]$ according to a uniform distribution. Then, 50000 sets of maps are  independently generated and split into training and validation sets with a ratio of 8:2. The training set is used to train MIMO-Unet model. Therefore, during training, our network can see CMB data with tensor-to-scalar ratios of $r\in [0.0001, 0.1]$. This means that the trained network model can handle unknown tensor-to-scalar ratios data in the range of $[0.0001, 0.1]$. To obtain the network predictions, we evaluate the trained MIMO-Unet model on three sets of 5000 test maps, and these three test sets have the fixed tensor-to-scalar ratio, $r_{\rm val} = 0.005$, $r_{\rm val} = 0.01$, $r_{\rm val} = 0.05$, respectively. Here, we choose these three fixed tensor-to-scalar ratio as an example to evaluate our network.

\begin{figure*}
\begin{center}
	\includegraphics[width=1\hsize]{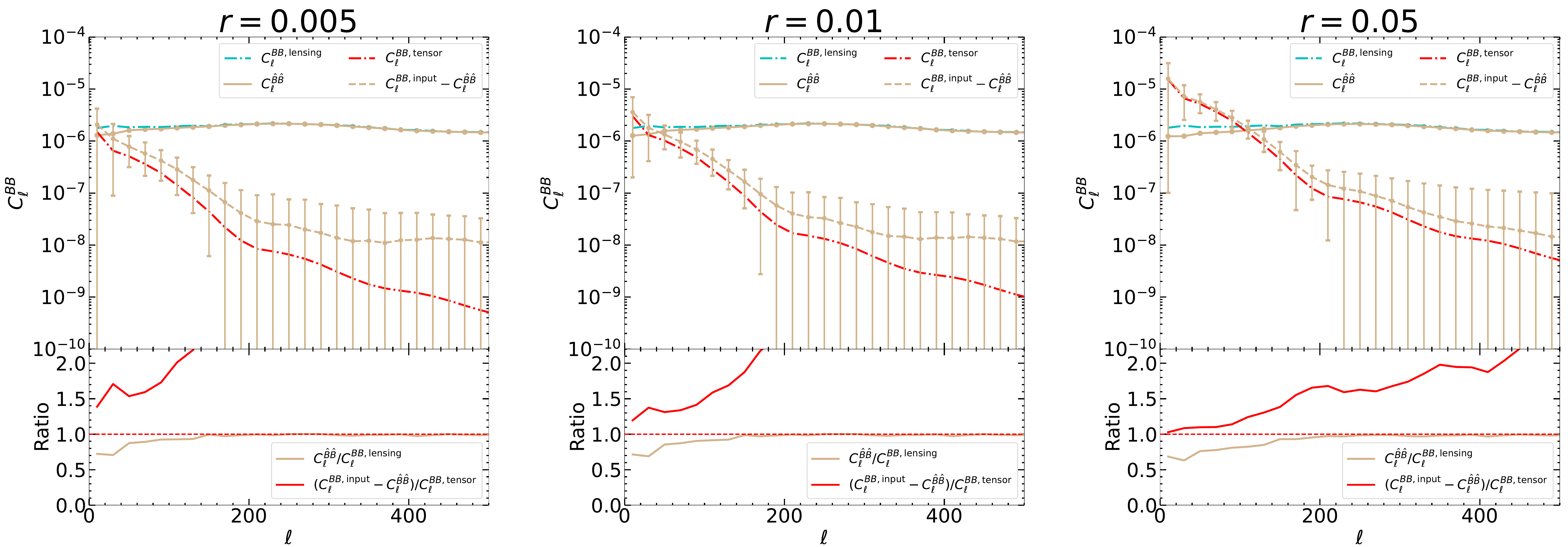}
\end{center}
\vspace{-0.4cm}
\caption{The recovered BB power spectra for experiments of three tensor-to-scalar ratio levels ($r=0.005,0.01,0.05$; left to right).  $C_{\ell}^{\rm BB,lensing}$ and $C_{\ell}^{\rm BB,tensor}$ are true lensing BB spectrum and primordial tensor BB spectrum. $C_{\ell}^{\hat{B}\hat{B}}$ is predicted lensing BB spectrum. $C_{\ell}^{\rm BB,input}$ is the network input BB power spectrum derived from observed CMB $Q$ and $U$ maps. We set the noise level to 1 $\mu$K-arcmin.}
\label{results_multi_BB_denoise_power}
\end{figure*}
We first apply the MIMO-Unet model to the task of delensing  CMB polarization maps similar to the Section \ref{delensing_result}. The observed CMB maps ($Q^{\rm observed}$, $U^{\rm observed}$) are used as input, and the desired output is delensing CMB maps ($Q^{\rm tar}$, $U^{\rm tar}$) with the instrument noise and beam. Figure \ref{results_multi_QU_denoise_power} plots the delensed BB spectra of three test sets after denoised step. As a discussion, we only show the 1 $\mu$K-arcmin noise level result; other noise level results are similar to this one. The BB spectra are averaged over the test sets. We see that the delens BB spectrum for the case of $r = 0.05$ is consistent with the tensor BB spectrum at the angular scales of $\ell<100$, which is worse than the result shown in Figure \ref{denoise_power} at $100<\ell<150$ where the network is trained on a fixed $r=0.05$ training set. Therefore, the variation of the tensor-to-scalar ratio on the training set will degrade slightly the reconstructed performance of the network at the angular scales of $100<\ell<150$. In addition, the reconstructed BB spectrum visibly degrades as $r$ decreases. The tensor B-mode power spectrum recovery at the angular scales of $\ell<100$ is about 77\% for observed maps with r = 0.01 and  56 \% for observed maps with r = 0.005. This implies that it is still very challenging for our network to reconstruct tensor BB spectrum with smaller $r$ through directly delensing Q and U maps. 

Next, we apply the MIMO-Unet model to the task of reconstructing lensing B map similar to the section \ref{Reconstruct_lensing}. The observed CMB maps ($Q^{\rm observed}$, $U^{\rm observed}$) are used as input, and the desired output is the observed lensing B map ($B^{\rm observed, lensing}$) with the instrument noise and beam. Figure \ref{results_multi_BB_denoise_power} shows the power spectra of reconstructed lensing B map for three test sets with $r = 0.005$,  $r = 0.01$, $r = 0.05$. We also only show the results of 1 $\mu$K-arcmin noise experiment. The reconstructed lensing BB power spectra are the mean result over the testing sets and are denoised by the cross-correlation of HS maps. For the case of $r=0.05$, we see that the reconstructed lensing BB power spectrum is quite consistent with the truth lensing BB spectra at $\ell>200$, and slightly deviates from the truth lensing BB at the angular scales of $\ell<200$. We also plot the delensing BB power spectrum using E.q. \ref{estimate}. We see that the delensing BB spectrum is basically consistent with the tensor BB spectrum at the angular scales of $\ell <200$. These results are consistent with the result shown in Figure \ref{results_BB_denoise_power} where the network is trained on a fixed $r=0.05$ training set. This implies that the variation of the tensor-to-scalar ratio on the training set will not significantly degrade the reconstructed lensing B map performance. In addition, our network can reconstruct lensing BB power spectrum for test sets with $r = 0.005$ and $r = 0.01$. However, delensing BB power spectrum using E.q. \ref{estimate} still degrades as $r$ decreases, which is consistent with the result above. The tensor B-mode power spectrum recovery at the angular scales of $\ell<100$ is around 71\% for observed maps with r = 0.01 and  56 \% for observed maps with r = 0.005.

\subsection{Variability of size of sky patch}
In section \ref{delensing_result} and \ref{Reconstruct_lensing}, we presented the results of delening CMB polarization maps and reconstructing lensing B map using the $20^{\circ} \times 20^{\circ}$ sized maps with $192 \times 192$ pixels.  Here, we investigate how the results depend on the size of sky patches. We simulate the flat maps that cover $15^{\circ} \times 15^{\circ}$ and $10^{\circ} \times 10^{\circ}$  patch of sky with $192 \times 192$ pixels as described in section \ref{data_pipe} and \ref{Reconstruct_lensing_results}. Considering that we simulated the flat maps, we do not increase the area of the sky area. As a discussion, we only show the results for 0 and 1 $\mu$K-arcmin noise level.

\begin{figure*}
	\begin{center}
		\includegraphics[width=1\hsize]{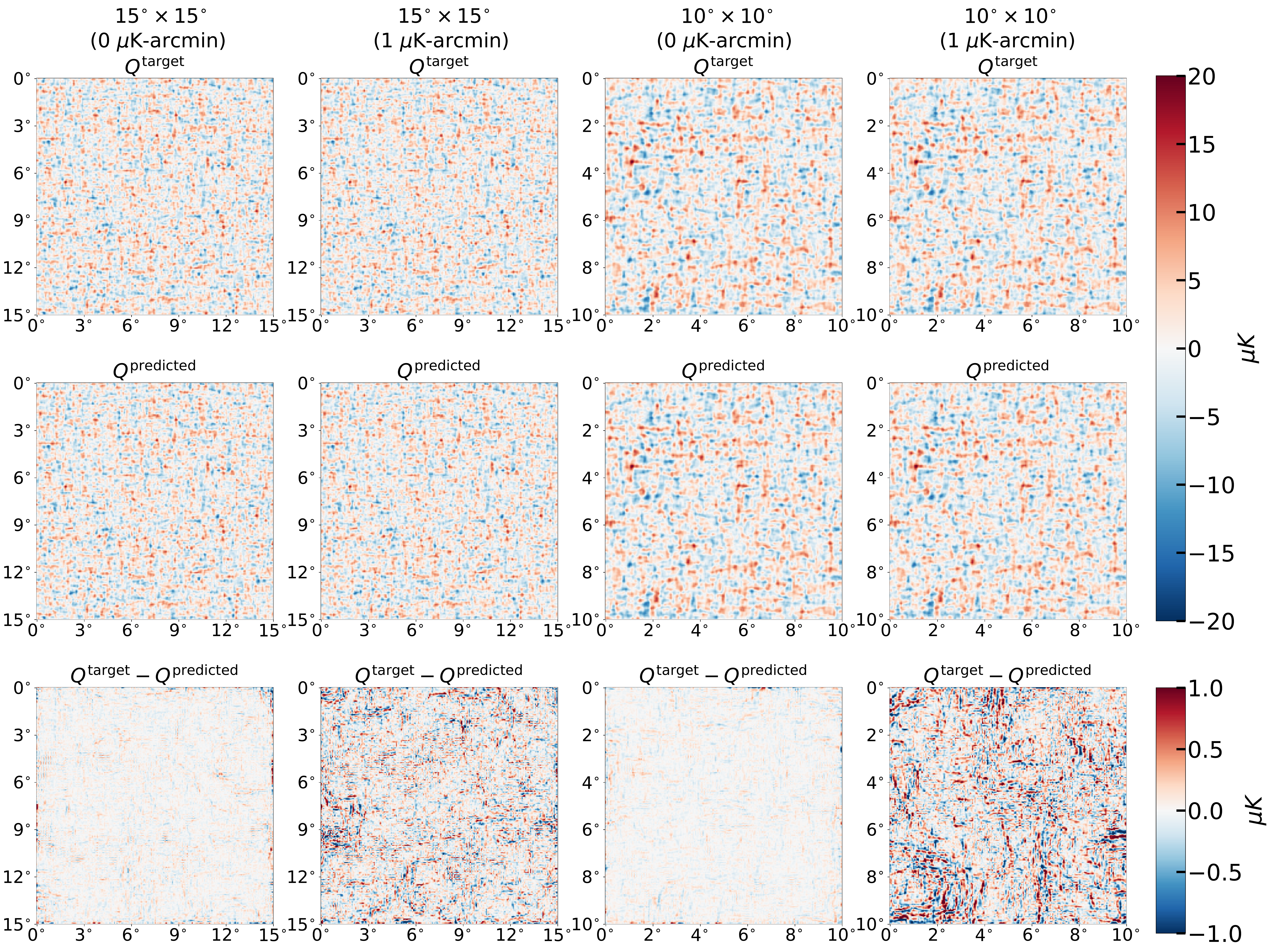}
	\end{center}
	\vspace{-0.4cm}
	\caption{Example of the delensing CMB Q  maps for  two different sized sky areas ($15^{\circ} \times 15^{\circ}$ and $10^{\circ} \times 10^{\circ}$)  and two different noise levels (0 and 1 $\mu$K-arcmin). The top row shows the target maps without lensing effect, but with noise and beam information. The middle row is the predicted delening maps by the MIMO-UNet. The bottom row shows the residual maps.}
	\label{results_map_patchs}
\end{figure*}

We first apply the MIMO-Unet model to the task of delensing CMB polarization maps similar to Section \ref{delensing_result}. The observed CMB maps ($Q^{\rm observed}$, $U^{\rm observed}$) are used as input, and the desired output is delensing CMB maps ($Q^{\rm tar}$, $U^{\rm tar}$) with the instrument noise and beam. Figure \ref{results_map_patchs} plots the delensed BB spectra for the experiments with two the sizes of sky patches ($15^{\circ} \times 15^{\circ}$ and $10^{\circ} \times 10^{\circ}$). In noiseless case, residual maps have minimal information left, implying that the MIMO-UNet can cleanly delens the Q maps for observed maps with patch sizes of $15^{\circ} \times 15^{\circ}$ and $10^{\circ} \times 10^{\circ}$. We calculate the average MAE between predicted delensing and target maps. As shown in table \ref{table_mae}, the average MAEs over test set from patch size of $15^{\circ} \times 15^{\circ}$ are $0.098 \pm 0.006\ \mu$K and $0.103 \pm 0.006\ \mu$K for delensing Q maps and U maps. The the average MAEs from patch size of $10^{\circ} \times 10^{\circ}$ are $0.075 \pm 0.004\ \mu$K and $0.068 \pm 0.004\ \mu$K for delensing Q maps and U maps. These values are slightly smaller than the results from sizes sky patches $20^{\circ} \times 20^{\circ}$ for the noiseless case. However, once the noise is present, the results are different from the noiseless case. As shown in table \ref{table_mae}, for the case of $1 \mu$K-arcmin noise, the average MAEs over the test set from patch size of $15^{\circ} \times 15^{\circ}$ are $0.274 \pm 0.014\ \mu$K and $0.286 \pm 0.013\ \mu$K for delensing Q maps and U maps. These values are consistent with the results from sizes sky patches $20^{\circ} \times 20^{\circ}$. The the average MAEs from patch size of $10^{\circ} \times 10^{\circ}$ are $0.331 \pm 0.020\ \mu$K and $0.330 \pm 0.019\ \mu$K for delensing Q maps and U maps. These values are slightly larger than the results from patch size of $20^{\circ} \times 20^{\circ}$. 

\begin{figure*}
	\begin{center}
		\includegraphics[width=1\hsize]{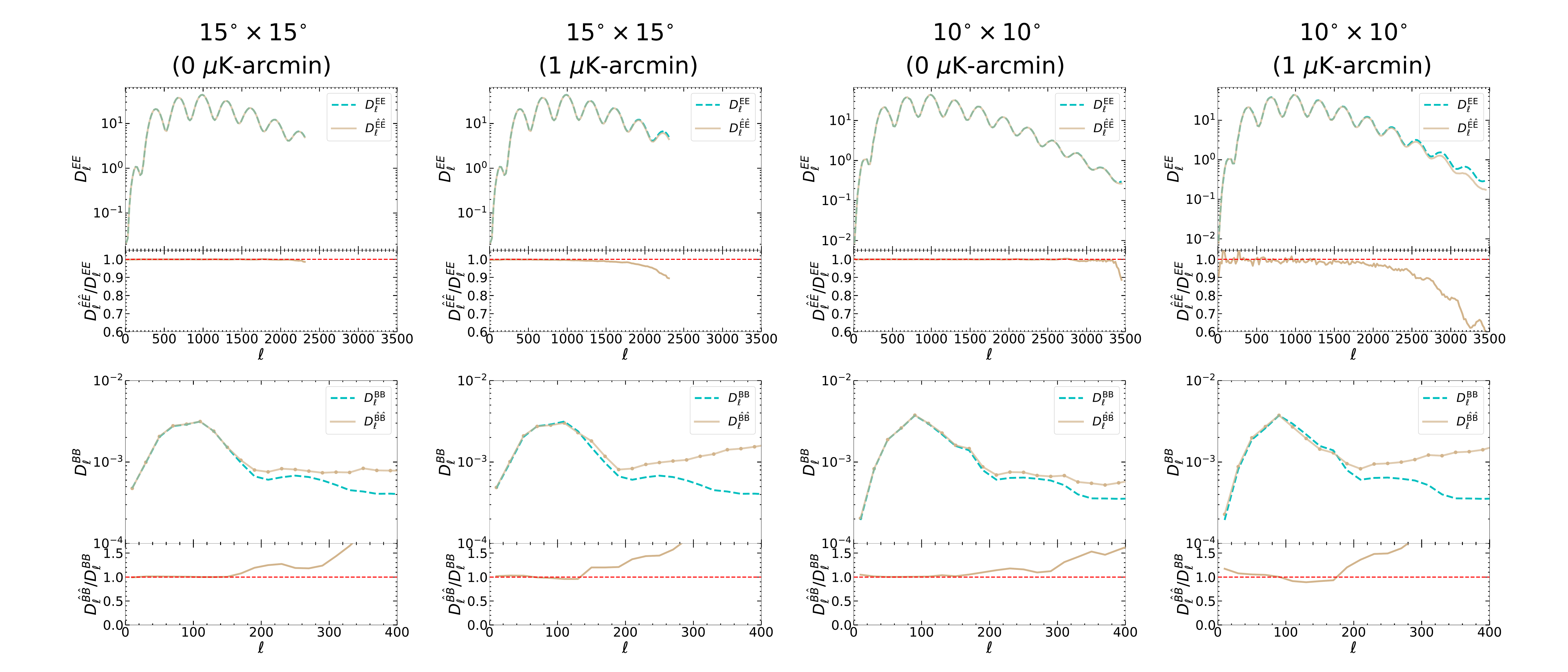}
	\end{center}
	\vspace{-0.4cm}
	\caption{The power spectra of the delensed EE (top panels) and BB (bottom panels) for two different sized sky areas ($15^{\circ} \times 15^{\circ}$ and $10^{\circ} \times 10^{\circ}$)  and two different noise levels (0 and 1 $\mu$K-arcmin). $\hat{E}$ and $\hat{B}$ represent predicted $E$ and $B$  obtained from our network's output ($Q, U$) maps.  $E$ and $B$ represent the target $E$ and $B$.  The length of each $\ell$ bin is set to be 20 here. }
	\label{results_denoise_power_patchs}
\end{figure*}

Figure \ref{denoise_power} shows the delensed EE and BB spectra after denoised step for the experiments with two the sizes of sky patches ($15^{\circ} \times 15^{\circ}$ and $10^{\circ} \times 10^{\circ}$).  The reconstructed $EE$ spectra ($D_{\ell}^{\hat{E}\hat{E}}$) are very consistent with primordial EE spectra ($D_{\ell}^{EE}$) at $\ell<3000$. The B-mode power spectrum recovers more than 98\% of the primordial B-mode at the angular scale of $\ell<150$, but it gradually deviates from the primordial BB ($D_{\ell}^{BB}$) as $\ell$ increases at the scales of $\ell>150$. These results are consistent with results of Section \ref{Delensing_results}. These imply that our results are independent of sky area size from the perspective of the recovered power spectrum.

Based on these analyses from recovered primordial (Q and U) maps and power spectra, we can conclude that the size of the sky area has a slight effect on the results, which is primarily seen in the recovered maps, but our network is still able to handle effectively data with a variety of sky patch sizes. Here, we do not show the dependence of the reconstructed lensing B-mode results (refer to Section \ref{Reconstruct_lensing_results}) on the size of sky patches, but we have checked that these conclusions drawn in this section still stand.

\begin{figure}
	\begin{center}
		\includegraphics[width=1\hsize]{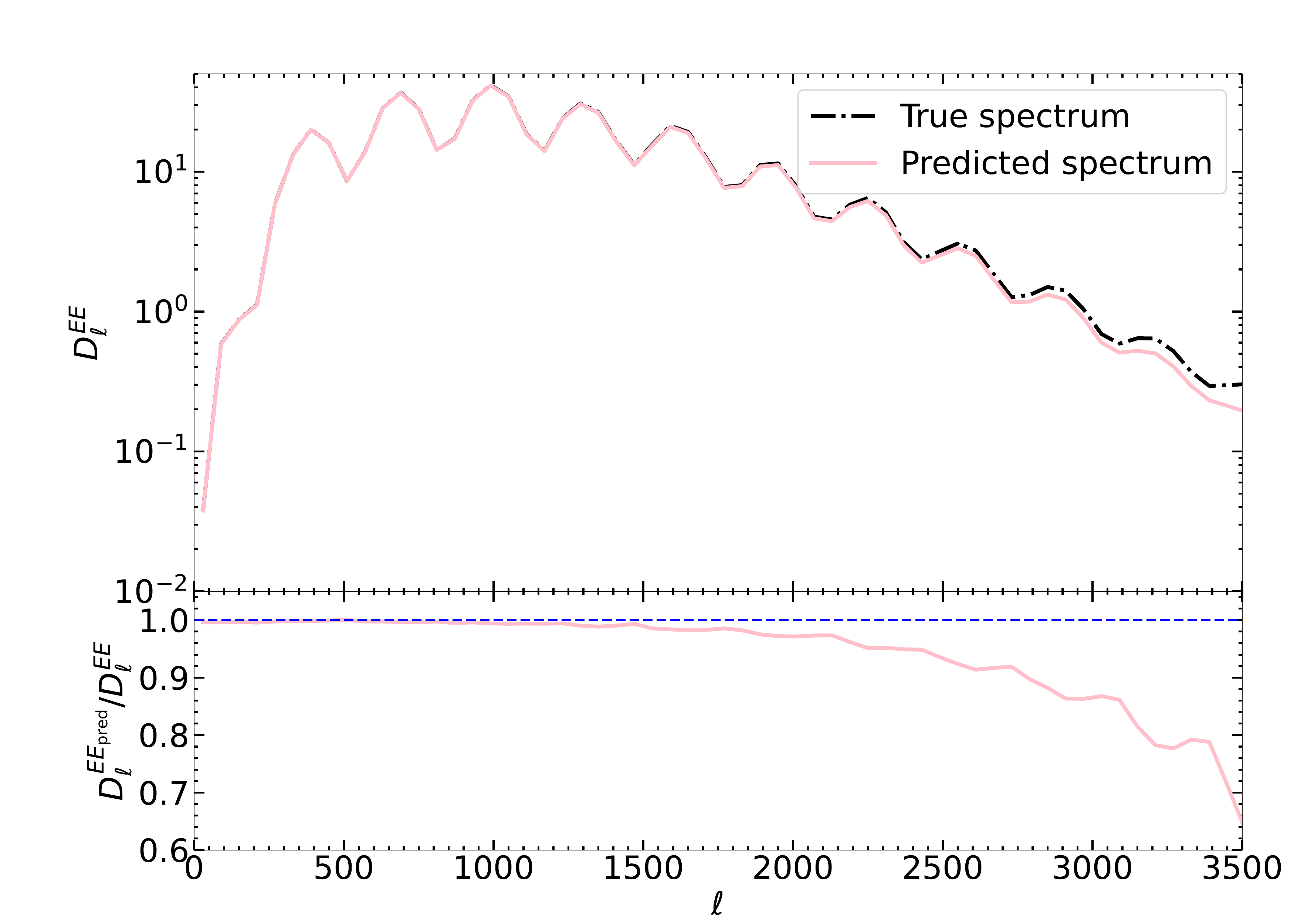}
	\end{center}
	\vspace{-0.4cm}
	\caption{The delensed EE power spectra for high-resolution CMB maps with $5^{\circ} \times 5^{\circ}$ patch of sky and $128 \times 128$ pixels. The true spectrum and predicted spectrum represent the primordial EE spectrum and delensed EE spectrum using the MIMO-UNet, respectively. We set noise level to 1 $\mu$K-arcmin.}
	\label{high_res_EE_power}
\end{figure}

We note that E-mode power spectrum recovery is greater than about 98\% for the angular scales of $\ell<2000$. At about $\ell =3000$, EE power spectrum recovery is around 80\%. This suggests that our network is very effective at recovering the primordial EE power spectrum. For comparison with the results of \cite{Guzman2021}, we simulate flat maps that cover $5^{\circ} \times 5^{\circ}$ a patch of sky with $128 \times 128$ pixels as described in section \ref{data_pipe}. This resolution of the map is consistent with \cite{Guzman2021}.  As a discussion, we only consider the case of  1 $\mu$K-arcmin noise level. The observed CMB maps ($Q^{\rm observed}$, $U^{\rm observed}$) are used as input, and the desired output is delensing CMB maps ($Q^{\rm tar}$, $U^{\rm tar}$) with the instrument noise and beam. We plot the delensed EE spectra after denoised step as shown in Figure \ref{high_res_EE_power}. We see that E-mode power spectrum recovery is greater than about 98\% for the angular scales of $\ell<2000$. At about $\ell =3000$, EE power spectrum recovery is around 87\%. This result is better than that of \cite{Guzman2021}. We do not show the B-mode power spectrum recovery because it is similar to the Section \ref{delensing_result} results.

\subsection{The CNN Method}
In this work, we demonstrated that the CNN can be used to recover unlensed CMB maps and lensing B map in simulated data. There are some challenges still present in the use of CNN method. The CNN method is to extract features from images and pass these features to later layers for the further filtering out of interesting signals. We find that our network is difficult to extract tensor B-model signals of the small scales because the features of tensor B-model map at small scale are very small. The large-scale features of the tensor B-model are challenging to recover when the tensor-to-scalar ratio is small $r<0.01$. In addition, the noise on the observed map will pollute the features of interesting signals and will degrade the performance of the network.  

\section{CONCLUSIONS}
\label{CONCLUSIONS}
In this paper, we presented a machine-learning CMB polarization delensing technique and demonstrated that the machine-learning can be used to delens CMB polarization maps in simulated data. The networks were trained using $20^{\circ} \times 20^{\circ}$ sized simulated CMB $Q$ and $U$ maps. We studied two delensing methods. For the first method, we use the network to reconstruct directly unlensing $Q$ and $U$ maps from observed $Q$ and $U$ maps. For the second method, we first use the network to reconstruct lensing B map from observed $Q$ and $U$ maps,  then delens observed BB power spectrum by subtracting this reconstructed lensing BB power spectrum. Note that our reconstructed maps for these two methods contain instrumental noise, and we use the cross-correlation technique to suppress the instrumental noise effect on the power spectrum. 

For the reconstruction of unlensing CMB $Q$ and $U$ maps, from map inspection,  our network  shows a good delening performance. The recovery of E-mode power spectrum is very consistent with the primordial EE spectrum across the entire range of angular scale we consider. The recovery of the B-mode power spectrum at noise levels of 0, 1, 2 $\mu$K-arcmin  is greater than about 98\% at the angular scales of $\ell<150$ but gradually deviates from the primordial BB as $\ell$ increases.

For reconstruction of lensing B map, our network can cleanly reconstruct lensing B map. The  recovery of the lensing B-mode power spectrum is greater than about 99\% at the angular scales of $\ell>200$ at noise levels of 0, 1, 2 $\mu$K-arcmin, slightly deviates from the truth lensing BB at the angular scales of $\ell<200$. We delens observed B-mode power spectrum by subtracting reconstructed lensing B-mode spectrum. The results show that the recovery of the tensor BB spectrum at noise levels of 0, 1, 2 $\mu$K-arcmin is greater than about 98\% at the scales of $\ell<120$. At about $\ell=160$, the recovery of tensor BB power spectrum is around 71\%.

We will further investigate the ability of our method to other aspects, such as the reconstruction of foregrounds and the component separation of future radio surveys, in future works

\section*{Acknowledgements}
J.-Q. Xia is supported by the National Science Foundation of China under grants No. U1931202 and 12021003; the National Key R\&D Program of China No. 2017YFA0402600 and 2020YFC2201603. 

\bibliography{bibliography_delensing.bib}
\bibliographystyle{aasjournal}

\end{document}